\documentclass[twocolumn,pra,amsmath,amssymb,amsfonts]{revtex4}
\usepackage{amssymb}
\usepackage{amsbsy}
\usepackage{epsf}
\usepackage{mathtext}
\usepackage{color}

\newcommand{\be}{\begin{equation}}
\newcommand{\ee}{\end{equation}}
\newcommand{\bea}{\begin{eqnarray}}
\newcommand{\eea}{\end{eqnarray}}

\usepackage[cp866nav]{inputenc}
\usepackage{revsymb}
\usepackage[T2A,OT1]{fontenc}
\usepackage{graphicx} 

\begin{document}

\title
{Dependence of the recoherence times and recoherence increments on the state of phonon bath in a single qubit dephasing model}
\author{V.V. Ignatyuk$^{a,b)}$, Ch. Samorodov$^{b)}$ 
}
\affiliation{a) Institute for Condensed Matter Physics of the National Academy of Sciences of Ukraine, 1
Svientsitskii Str., 79011, Lviv, Ukraine\\
b) Ivan Franko National University of Lviv, 1 Universytetska Str., Lviv, 79007, Ukraine
}

\date{\today}

\begin{abstract}
\noindent
The recoherence times $t^*$ and the maximum values of the recoherence increments $\gamma_{\rm extr}$ are studied as functions
of the bath parameters for a single qubit dephasing model, prepared initially by a special kind of the non-selective measurements. The recoherence/decoherence events (RDE), occurring at the initial stage of the system evolution, are found to be both similar and different from the system dynamics at large times. For instance, in contrast to the RDE observed on large time scales, the sub-Ohmic and Ohmic coupling regimes are more favourable for the short-time recoherence than the super-Ohmic one. On the other hand, the short-time behaviour of the recoherence and the long-time dynamics of the decoherence are closely related: the domain of the ohmicity indexes, where the decoherence changes its type (from the complete to incomplete one), is, simultaneously, that of the weakest recoherence. The obtained results give us some hints about the basic characteristics of the environment, which might provide the most optimal values of $t^*$ and $\gamma_{\rm extr}$ in some sense.

\end{abstract}

\maketitle

\section{Introduction\label{Introduct}}

The quantum coherence (QC) along with the superposition of states and the entanglement is one of the most fascinating phenomena of quantum nature, which have been attracting a close attention of scientists for a long time \cite{Wineland}. Usually, when talking about QC, we mean two opposite tendencies in its manifestation. The unavoidable environmentally induced \textit{decoherence} phenomenon is reasonably believed to play a major destructive role in the dynamics of two-state systems of qubits, which are the elementary carriers of quantum information~\cite{q-measurement1,39min}.  

On the other hand, the \textit{recoherence} phenomenon (also referred to as the coherence enhancement \cite{myPRA2015} or coherence trapping \cite{trapping}) is another pathway of the QC evolution, generally of a constructive nature. It occurs when a decay of QC is terminated at a certain instant, and the system can partly retain coherence for a finite time period or in the long time limit,  or even revive it. This manifestation of the QC duality was thoroughly studied for a variety of quantum systems such as the charge donor qubits \cite{charge-q-bit}, the electrons coupled to the quantized electromagnetic field 
\cite{Hsiang2009}, in the pigment-protein complexes used in the light harvesting technologies 
\cite{pp-complexes}, in the quantum computer simulations \cite{chisholm2021witnessing}, and many others. The recoherence also underlies the quantum error correction and the related techniques \cite{macchiavello2000quantum}, thus having a significant practical application in quantum computing.

The recoherence is often studied from the viewpoint of the memory effects in the open quantum systems \cite{reina2002,haikka2013non,nonMark-1}, invoking the notion of the so-called non-Markovianity \cite{budini,lombardo2015}. The non-Markovian dynamics implies an information flow from the environment back to the open quantum system \cite{giraldi2017}, yielding a growth of the trace distance between two initial quantum states and increasing their relative distinguishability \cite{clos2012}. Here the keyword is \textit{the information backflow}, since in many cases (including the studies of the present paper), an energy exchange between the quantum open system and its environment is absent, and one can talk only about the entropy exchange expressed in terms of the Bloch vector length \cite{PRA2012}. Within the master equation approach, it corresponds to the emergence of negative values of the dephasing rates \cite{gamma-minus-1,gamma-minus-2,gamma-minus-3,colloquium2016,rivas2014}, which, in its turn, is directly related to the above mentioned non-Markovian dynamics.

The recoherence is widely recognized to depend on the nature of the qubit-environment interaction \cite{haikka2013non,lombardo2015,giraldi2017}. It becomes noticeable if the ohmicity index $s$ exceeds the value~2. Within the dephasing model \cite{PRA2012}, it corresponds to the incomplete (partial) decoherence at the long times $t\to\infty$. Usually, a typical hierarchy in the ohmicity indexes is observed: the bigger is the value of $s$, the more pronounced is the recoherence induced by the environment \cite{lombardo2015}. In a general case, a magnitude of the recoherence depends also on the bath temperature $T$. For instance, at high temperatures, the threshold value of $s$ of the recoherence emergence increases up to $s=3$. Even more remarkable phenomenon was reported in Ref.~\cite{giraldi2017}, where the decoherence/recoherence processes had been found to be governed both by the shape of the spectral weight function and by temperature. Therefore, a possibility to control the system coherence and to switch the direction of information fluxes at different times appears, offering a huge prospect for application in quantum information technologies. 

All the above mentioned conclusions concern  the recoherence dynamics, when the open quantum system and its environment were chosen to be uncorrelated at the initial instant of system evolution, while the coherence enhancement was measured at large times. It means that at $t=0$, the density matrix of the total system $\varrho(0)$ is chosen as a direct product, $\varrho(0)=\varrho_S\otimes\varrho_B$, of those of the open quantum (sub)system $\varrho_S$ and the environment (the thermal bath) $\varrho_B$. However, back in the early 2000-ies a much more interesting behaviour of the recoherence in the chain of quantum registers was reported \cite{reina2002} with more general initial conditions, where the initial state of the composite system is allowed to contain some correlations between the surrounding environment and the qubits. As for the dephasing model used in the present paper, the correlated initial states were studied in Ref.~\cite{luczka2013} in the context of the system purification, which is a direct consequence of the coherence enhancement. 

These studies were continued in Ref.~\cite{myPRA2015} by selection of a more physical initial conditions, imposed on the composite ($S+B$) system by \textit{the non-selective measurements} \cite{BPbook}. In~\cite{myPRA2015}, there was observed a sequence of RDE occurring at the initial stage of the system evolution. A special kind of the non-selective measurements has been  proposed, which ensures an indubitable coherence enhancement regardless of the qubit state vectors, as long as the ``qubit-bath'' interaction strength is large enough. Obviously, at larger times the vacuum and thermal fluctuations will overcome the impact of initial correlations, and one comes back to the cases described above, when RDE are possible only in the advanced super-Ohmic coupling regimes with $s>2$. It is also worthwhile to mention that the initial qubit-bath correlations can also suppress the QC \cite{PRA2012}, as happens at \textit{the selective measurements} \cite{BPbook}, when an additional channel of the decoherence appears.

Though in most cases the recoherence processes are studied at long times, when the impact of initial correlations has already finished, the short-time dynamics of RDE is also important if one deals e.~g. with a dynamic decoupling technique to freeze the quantum evolution of the system \cite{chaudhry2013} or performs the repeated measurements on it \cite{overcomplete}. The short-time RDE of a qubit subjected to the non-selective measurements were studied in detail in Refs.~\cite{giraldi2021,giraldi2022short} by the time series expansion that allowed also to count the total number of RDE as the function of coupling strength, ohmicity index, and bath temperature. However, even though one deals with a comparatively short times (as compared to the inverse rate of decoherence) when studying an influence of the correlation contribution on the coherence enhancement \cite{myPRA2015,overcomplete}, a total period of the QC growth can be large enough to use a reliable truncation of the above series. Therefore, no analytical expression for the dephasing rates (as those derived in \cite{giraldi2022short}) can be obtained, and only numerical calculation of the recoherence times and  increments is plausible. At the same time, a study of dependence of the recoherence times and increments, caused by the initial correlations, on the bath characteristics remains a relevant problem in many cases.

In this paper, the recoherence times $t^*$ and the maximum values of the recoherence increments $\gamma_{\rm extr}$ are studied as functions of coupling strengths, ohmicity indexes, and bath temperature. We found that in contrast to the RDE observed at long times, the sub-Ohmic ($s<1$) and Ohmic ($s=1$) regimes are more favourable for the recoherence than the super-Ohmic
one ($s>1$). The most interesting observation is that the short-time behaviour of the recoherence and the long-time dynamics of the decoherence can be interrelated: the domain $s\ge 2$, where the decoherence changes its type (from the complete to incomplete one \cite{PRA2012}), is, simultaneously, that of the weakest recoherence. The obtained results give us some insights about the basic characteristics of the environment, which can provide the most optimal values of $t^*$ and $\gamma_{\rm extr}$ for an application in the quantum measurement experiments.

The paper is structured as follows. In Section~\ref{secII}, we recall the basic relations corresponding to the non-selective quantum measurements. A special attention is paid to the measurement states, which provide a diagonal form of the Gram operator \cite{myPRA2015,overcomplete}. In Section~\ref{secIII}, starting from the Hamiltonian of the one qubit dephasing model, we present the expressions for the generalized decoherence function $\gamma_{\rm tot}(t)$ and its constituent $\gamma_{\rm cor}(t)$ related to initial correlations in the composite system. The latter expression is simplified for the case of the diagonal Gram operator. In Section~\ref{secIV}, using this simplified version of $\gamma_{\rm cor}(t)$, we perform the numerical calculation of the recoherence times and the recoherence increments depending on the bath parameters : the coupling strength, the ohmicity index, and the temperature. The obtained results are discuss in detail in their relation to a critical regime of the recoherence emergence. In the last Section, we conclude our studies and outline some problems for future investigation.

\section{Qubit initial state preparation by the non-selective measurements: a detailed examination\label{secII}}

Suppose that at all times $t<0$ an open system $S$ was in the thermal
equilibrium with a heat bath $B$, and at time zero one makes a
measurement on the system $S$ only. According to the general
principles of quantum measurement theory
\cite{BPbook,q-measurement,kraus}, the initial state of the composite
system ($S+B $) after the measurement is described by the
density matrix
 \begin{equation}
    \label{rho-SB-0}
 \varrho^{}_{SB}(0)= \sum_{m}\Omega^{}_{m}\varrho^{}_{\text{eq}}\Omega^{\dagger}_{m},
 \end{equation}
where $\varrho^{}_{\text{eq}}$ means the equilibrium density matrix at temperature $T$. Operators $\Omega^{}_{m}$ act in the Hilbert space of the system $S$ and correspond to
 possible outcomes $m$ of the measurement.
In a particular case of the selective measurement, the system $S$ is prepared in some
pure state $|\psi\rangle $. Then the sum in Eq.~(\ref{rho-SB-0}) contains only one
term, so that
  \begin{equation}
    \label{rho-SB-selec}
     \varrho^{}_{SB}(0)=
     P^{}_{\psi}\varrho^{}_{\text{eq}}P^{}_{\psi},
  \end{equation}
where $ P^{}_{\psi}=|\psi\rangle \langle\psi| $ is the projector onto the quantum state $|\psi\rangle$.

In general, the density matrix (\ref{rho-SB-0}) describes the resulting
ensemble after the non-selective measurement, in which
the outcome $m$  may be viewed as a classical random number with the
probability distribution
    \begin{equation}
   \label{p-m}
  w(m)=\text{Tr}\left\{F^{}_{m}\varrho^{}_{\rm eq}\right\}.
  \end{equation}
The positive hermitian operators $F^{}_{m}=\Omega^{\dagger}_{m}\Omega^{}_{m}$ are called the ``effects'' \cite{myPRA2015,BPbook,q-measurement,overcomplete}.
   Hereafter, $\text{Tr}$ denotes the trace over the Hilbert space of the
    composite ($S+B $) system, while the symbols $\text{Tr}^{}_{S}$ and $\text{Tr}^{}_{B}$
     will be used to denote the partial traces over the Hilbert spaces of the system
     $S$ and the heat bath, respectively. 
     
     The measurement operators $\Omega_m$ are defined \cite{myPRA2015,overcomplete} as 
     \begin{equation}
     \label{Omega-m-gen}
     \Omega^{}_{m}=|\varphi^{}_{m}\rangle \langle \psi^{}_{m}|.
         \end{equation}
A precise form of $\Omega^{}_{m}$ is determined by the details of the measuring device \cite{BPbook}.
From a mathematical viewpoint, the values (\ref{Omega-m-gen}) can be treated as some kind of the ``skew-projecting'' operators defined on the pure states $|\varphi^{}_{m}\rangle $ and $|\psi_{m}\rangle$. Obviously, these sets of the measurement states can be related to each other via some unitary transformations
       \begin{equation}
     \label{phi-m}
     |\varphi^{}_{m}\rangle=U^{}_{m} |\psi^{}_{m}\rangle\, .
     \end{equation}
 Hence the non-selective measurement scheme (\ref{rho-SB-0}) can be considered as a combined effect of the selective (projective) measurements (\ref{rho-SB-selec}), the unitary transformations (\ref{phi-m}), and the subsequent mixing of the obtained quantum states with the probabilities (\ref{p-m}).

  In order to the initial density matrix (\ref{rho-SB-0}) be normalized to unity, 
 \begin{equation}
 \text{Tr}\varrho^{}_{SB}(0)\!
 =\!
 \sum_{m} \text{Tr}\{\varrho^{}_{\text{eq}}\Omega^{\dagger}_{m}\Omega^{}_{m}\}=1,
 \end{equation} 
  the effects have to obey the following relation:
    \begin{equation}
    \label{Fm-NormCond}
  \sum_{m}F^{}_{m}\equiv \sum_{m}\Omega^{\dagger}_{m}\Omega^{}_{m}=I,
  \end{equation}
where $I$ means the identity operator.

Some essential remarks are in position. Whereas the states $|\psi^{}_{m}\rangle$ form an
   orthonormal complete basis according to (\ref{Fm-NormCond}), this condition holds true for the transformed states  $|\varphi^{}_{m}\rangle$
   only if the unitary operators $U^{}_{m}$  are identical for all
  outcomes $m$ (i.~e., $U^{}_{m}=U$).
    In such cases, taking into account that $\Omega^{}_{m}\Omega^{\dagger}_{m}=|\varphi^{}_{m}\rangle\langle\varphi^{}_{m}|$ and introducing the Gram operator $G=\sum_{m}\Omega^{}_{m}\Omega^{\dagger}_{m}$ \cite{myPRA2015,overcomplete}, we
     have another resolution of the identity in addition to Eq.~(\ref{Fm-NormCond}),
      \begin{equation}
        \label{Phi-norm-cond}
     G= \sum_{m}\Omega^{}_{m}\Omega^{\dagger}_{m}=I\,.
      \end{equation}
      
 If the form of $U^{}_{m}$ depends on the outcome $m$, then
 the transformed states $|\varphi^{}_{m}\rangle$
 are not orthogonal in general. Some of these states may even be identical \cite{myPRA2015}. In our research, we will consider the special measurement schemes, when the corresponding Gram operator either satisfy condition (\ref{Phi-norm-cond}) or is constructed to be diagonal, $G=\mbox{diag}$.

Another remark is about the interrelations between the number $m=\{1,\ldots,N\}$ of measurements and the rank $n$ of the reduced density matrix (for qubits, $n=2$). If $n=N$, all the effects $F_m$ are projectors ($F^{2}_{m}=F^{}_{m}$). This is a natural generalization of the well-known von Neumann-L\"uders projection postulate for ideal quantum measurements
(see, e.g., Ref.~\cite{BPbook}). One can also construct more general measurement
schemes associated with the notion of the positive operator-valued measure (POVM)~\cite{Holevo2001,SRM}. A POVM is defined by $N$ positive operators $F^{}_{m}$, which form the so-called non-orthogonal resolution of the identity, but in general $F^{2}_{m}\not=F^{}_{m}$.
Usually, it occurs when the basic sets of measurement vectors $\{|\psi^{}_{m}\rangle\}$, $\{|\varphi^{}_{m}\rangle\}$ are overcomplete, i.e., the number of outcomes $N$ exceeds the rank $n$ of the density matrix of the system. However, even in this more general case, when the measurement state vectors can no longer be orthonormal, one can construct the diagonal Gram operator, which ensures the enhancement of qubit coherence at the initial period of the open system evolution \cite{overcomplete}.

 Let us now apply the above general construction to a qubit.
 In the formal ``spin'' representation,
 the canonical orthonormal basis states of a qubit are
  \begin{equation}
    \label{qbit-can-bas}
  |0\rangle= \left( \begin{array}{c} 0\\ 1 \end{array}\right),
  \quad
  |1\rangle= \left(\begin{array}{c} 1\\ 0 \end{array}\right).
  \end{equation}
All pure states
  $|\psi(\vec{a})\rangle\equiv |\vec{a}\rangle $ correspond to the points on the Bloch sphere
   $|\vec{a}|$ = 1, where $\vec{a}=\left(a^{}_{1},a^{}_{2},a^{}_{3}\right)\in \mathbb{R}^{3}$.
The normalized state vectors are given by (see, e.g.,~\cite{Holevo2001})
\begin{equation}
\label{a-qubit}
|\vec{a}\rangle=
\left(
\begin{array}{c}
\displaystyle {e}^{i\phi_a/2}\,\sin (\theta_a/2)\,\\[7pt]
\displaystyle {e}^{-i\phi_a/2}\,\cos(\theta_a/2)
\end{array}
\right),
\end{equation}
     where $\phi_a$ and $\theta_a$ are the Euler angles of the unit vector $\vec{a}$ describing the direction of the ``spin''. They satisfy the relations
    $a^{}_{1}+ia^{}_{2}= \sin\theta_a\, {e}^{i\phi_a}$, $a^{}_{3}= \cos\theta_a$.
     The Euler angles corresponding to the state $|-\vec{a}\rangle$, which is orthogonal to (\ref{a-qubit}), are
      \begin{equation}
       \label{Euler;-a}
       \theta^{}_{-a}=\pi - \theta^{}_{a}, \quad \phi^{}_{-a}=\phi^{}_{a} +\pi.
      \end{equation}
Using these relations together with Eq.~(\ref{a-qubit}) gives
 \begin{equation}
\label{-a-qubit}
|-\vec{a}\rangle=
\left(
\begin{array}{c}
\displaystyle i{e}^{i\phi_a/2}\,\cos(\theta_a/2)\,\\[7pt]
\displaystyle -i {e}^{-i\phi_a/2}\,\sin(\theta_a/2)
\end{array}
\right).
\end{equation}

    All operators in the qubit's Hilbert space can be represented as linear combinations of the unity operator $I$  and the Pauli matrices $\sigma^{}_{1},\sigma^{}_{2},\sigma^{}_{3}$. For example,  the operator
      \begin{equation}
        \label{sigma-a}
       \sigma(\vec{a})=\sigma^{}_{1}a^{}_{1} + \sigma^{}_{2}a^{}_{2}+\sigma^{}_{3}a^{}_{3}
      \end{equation}
  describes the spin component in the direction $\vec{a}$. The state vectors
  $|\pm\vec{a}\rangle$ correspond to the eigenvalues $\pm 1$ of $\sigma(\vec{a})$
   and form an orthonormal basis,
     \begin{equation}
       \label{a-a-bas}
     |\vec{a}\rangle \langle\vec{a}|+  |-\vec{a}\rangle \langle-\vec{a}|=I.
     \end{equation}

The selective measurement \cite{PRA2012} applied to a qubit can be characterized by the
projector $P(\vec{a})= |\vec{a}\rangle \langle\vec{a}|$,
 while the general non-selective measurement scheme is associated with three states: $|\psi_1\rangle=|\vec{a}\rangle$,  $|\varphi_i\rangle=|\vec{b}^{}_{i}\rangle$, $i=\{1,2\}$, since the fourth state vector has always to be chosen as $|\psi_2\rangle=|-\vec{a}\rangle$.

Let us briefly consider some important special cases of the non-selective measurement scheme:

 i)  $\vec{b}^{}_{1}= \vec{a},\ \vec{b}^{}_{2}=- \vec{a}$. This is the
  simplest scheme corresponding to
  $U^{}_{m}=I$ in the general formula (\ref{phi-m}). Physically, here we are dealing
   with the non-selective measurements, where the measuring  device does not disturb the
    basis states $|\vec{a}\rangle$ and $|-\vec{a}\rangle$. In this case, the
     $\Omega$-operators coincide with the effects:
       \begin{equation}
       \label{Uunity}
    \Omega_1=F_1=|\vec a\rangle\langle\vec a|, \qquad
    \Omega_2=F_2=|-\vec a\rangle\langle-\vec a|\, .
      \end{equation}

ii) $\vec{b}^{}_{1}=\vec{b}$,\ $\vec{b}^{}_{2}=- \vec{b}$, where $\vec{b}$ is an
 arbitrary unit vector. This case corresponds to $U^{}_{m}=U\equiv U(\vec{b},\vec{a})$ in
Eq.~(\ref{phi-m}). The explicit form of the unitary operator $U(\vec{b},\vec{a})$ can be found for each particular choice of the state vectors $|\vec{a}\rangle$ and $|\vec{b}\rangle$. The Euler angles $\theta_i$, $\phi_i$ of the pair of two ortho-normalized vectors  $|\vec{b}^{}_{i}\rangle$ can be shown to satisfy the relations
 \begin{equation}\label{cond-angles-1}
 \theta_1+\theta_2=\pi,\qquad  \Delta_{\phi}=\phi_1-\phi_2=\pi.
 \end{equation}
 It means that (like for the pair of state vectors $|\vec{a}\rangle$ and $|- \vec{a}\rangle$) one actually needs to introduce only two Euler angles to specify the measurement scheme. 

iii) $\vec{b}_{1}\ne-\vec{b}_2$, but the Gram operator is diagonal, $G=\sum_i |\vec{b}_{i}\rangle\langle \vec{b}_{i}|=\mbox{diag}$. In Ref.~\cite{myPRA2015}, the pair of two quantum states was considered,
\begin{eqnarray}
\label{b12-ampl}
|\vec{b}^{}_{1}\rangle= c^{}_{0} |0\rangle + c^{}_{1} |1\rangle,\qquad
|\vec{b}^{}_{2}\rangle= i\big( c^{}_{0} |0\rangle - c^{}_{1} |1\rangle \big),
\end{eqnarray}
where the amplitudes $ c^{}_{0}$ and $ c^{}_{1}$ can be expressed in terms
of the Euler angles by using Eq.~(\ref{a-qubit}). It can be easily shown, that the corresponding Euler angles obey the following relations (cf. Eq.~(\ref{cond-angles-1})):
\begin{equation}\label{cond-angles-2}
\theta_1=\theta_2,\qquad  \Delta_{\phi}=\phi_1-\phi_2=\pi.
\end{equation}
It is straightforward to verify that for the basic set of the measurement vectors (\ref{b12-ampl}), the corresponding Gram operators reads as
\begin{eqnarray}
G=2\left(
\begin{array}{cc}
|c_1|^2 & 0\\
0 & |c_0|^2
\end{array}
\right),
\end{eqnarray}
For the unitary operators, relating the sets $|\psi_i \rangle$ and $|\varphi_i \rangle$, a partial solution can be found as
\begin{eqnarray}\label{U1diag}
U_1=I,\qquad U_2=\left(
\begin{array}{cc}
0 & e^{i\phi_a}\\
e^{-i\phi_a} & 0
\end{array}
\right).
\end{eqnarray}
In a similar way to (\ref{b12-ampl}), one can consider another pair of the state vectors $|\varphi_i\rangle$ forming the diagonal Gram operator, which are related to the basic set $|\psi_i\rangle$ via the unitary transformations
\begin{eqnarray}\label{U2diag}
U_1=\sigma_3,\qquad U_2=\left(
\begin{array}{cc}
0 & i e^{i\phi_a}\\
-i e^{-i\phi_a} & 0
\end{array}
\right).
\end{eqnarray}
It is seen from Eqs.~(\ref{U1diag})-(\ref{U2diag}) that at the real amplitudes $c_i$, the unitary transformations are constructed on the basis of the identity matrix and the Dirac operators $\sigma_i$, $i=\{1,2,3\}$.
In the general case, the explicit form of the unitary operators $U_i$, yielding the diagonal Gram matrix $G=\mbox{diag}$, can be derived from the following relation:
\begin{equation}\label{Udiag}
U_1|\psi_1\rangle\langle\psi_1|U_1^{\dagger}-U_2|\psi_1\rangle\langle\psi_1|U_2^{\dagger}=G-I,
\end{equation} 
where $|\psi_1\rangle$ means an arbitrary pure quantum state.

\section{The dephasing model: dynamics of decoherence at the non-selective measurements\label{secIII} }
 
In this Section, we briefly analyse the simple \textit{dephasing model\/}
 describing the main decoherence mechanism
 for certain types of system-environment  interactions~\cite{PRA2012,luczka2013,Unruh,MR-CMP2012,myCMP}.
 In this model, the two-state system (qubit) ($S$) is coupled to a bath ($B$) of
harmonic oscillators. Using the ``spin'' representation for the qubit with the basis states
 (\ref{qbit-can-bas}), the total Hamiltonian is taken to be
 (in our units $\hbar = 1$)
    \begin{eqnarray}
    \label{H}
    \nonumber
  &&  \hspace*{-20pt} H=H_S+H_B+H_{int}\\
   && \hspace*{-20pt} {}=\frac{\omega_0}{2}\sigma_3
   +\sum\limits_k\omega_k b^{\dagger}_k b_k
   +\sigma_3\sum\limits_k(g_k b^{\dagger}_k+g^*_k b_k),
   \end{eqnarray}
where $\omega_0$ is the energy difference between the excited
 state $|1\rangle$ and the ground state $|0\rangle$  of the qubit.
Bosonic creation and annihilation operators $b^{\dagger}_k$ and
$b_k$ correspond to the $k$th bath mode with frequency $\omega_k$,
and $g_k$ are the coupling constants.

Suppose that at time $t=0$ the state of the composite system ($S$+$B$) is characterized
by some density matrix $\varrho^{}_{SB}(0)$. Then at time $t$, the average value
 of a Heisenberg picture operator $A(t)$ is given by
 \begin{equation}
    \label{A(t)}
   \langle A(t)\rangle= \text{Tr}\left\{
     \exp(iHt)A\exp(-iHt)\varrho^{}_{SB}(0)
   \right\}.
  \end{equation}
Hereafter, the notation $\langle A\rangle$ is used  for the averages at $t = 0$.

 The quantities of principal interest are the
  \textit{coherences\/} $\langle \sigma^{}_{\pm}(t)\rangle$, where
   $\sigma^{}_{\pm}=\left(\sigma^{}_{1}\pm i\sigma^{}_{2}\right)/2$.
  They are related directly to
   the off-diagonal elements of the reduced density matrix of the qubit:
    \begin{equation}
       \label{coher-DM}
      \langle \sigma^{}_{+}(t)\rangle=\langle 0| \varrho^{}_{S}(t)|1\rangle,
      \qquad
       \langle \sigma^{}_{-}(t)\rangle= \langle 1| \varrho^{}_{S}(t)|0\rangle,
    \end{equation}
 where
   \begin{equation}
      \label{rho-S}
     \varrho^{}_{S}(t)= \text{Tr}^{}_{B}
     \left\{
          \exp(-iHt) \varrho^{}_{SB}(0)\exp(iHt)
     \right\}.
   \end{equation}

Within the dephasing model (\ref{H}), the equations of motion for
  all relevant operators are solvable exactly~\cite{PRA2012}. This allows one to
   study the time evolution of the coherences for different initial preparation
   conditions. We leave out some less essential details for which we refer the reader to
    Ref.~\cite{PRA2012}, and quote the major results.

    If the initial state is prepared by the non-selective measurement, then,
    taking the initial density matrix of the composite system in the form
       (\ref{rho-SB-0}), we get  the following expression for the coherences (\ref{coher-DM})
       \begin{equation}
    \label{coh1}
   \langle\sigma_{\pm}(t)\rangle=\frac{1}{Z}\sum_{m}\text{Tr}
   \left[\Omega_m^{\dagger}\sigma_{\pm}(t)\Omega_m e^{-\beta H}\right],
  \end{equation}
  where $\beta=1/T $ (we put the Boltzmann constant to be equal to 1), and $Z=\text{Tr}\left\{\exp(-\beta H)\right\}$
  is the equilibrium partition function.

    As shown in Ref.~\cite{PRA2012}, the time-dependent
    qubit operators $\sigma^{}_{\pm}(t)$ in the dephasing model (\ref{H}) are given by
       \begin{equation}
       \label{sig-pm-t}
       \sigma_{\pm}(t)= \exp\left\{ \pm i\omega_0 t \mp R(t) \right\} \sigma_{\pm}
       \end{equation}
   with
      \begin{equation}
    \label{R_alpha}
    R(t)=\sum_{k}\!\left[\alpha_{k}(t) b^{\dagger}_k-\alpha^*_k(t) b_{k} \right],
  \quad
  \alpha_k(t)\!=2g_{k}\frac{1-e^{i\omega_kt}}{\omega_{k}}.
 \end{equation}
Using the above expressions and the exact relations
   \begin{equation}
   \label{rho_relations}
   \begin{array}{l}
    e^{-\beta H}|0\rangle= e^{\beta\omega_0/2}e^{-\beta H^{(-)}_B}\otimes|0\rangle,\\[5pt]
     e^{-\beta H}|1\rangle= e^{-\beta\omega_0/2}e^{-\beta H^{(+)}_B}\otimes|1\rangle,
   \end{array}
   \end{equation}
where
  \begin{equation}
  \label{HBpm}
  H_B^{(\pm)}=\sum\limits_k\omega_k b^{\dagger}_k  b_k
  \pm\sum\limits_k(g_k b^{\dagger}_k+g^*_k b_k),
 \end{equation}
the trace over the bath degrees of freedom in Eq.~(\ref{coh1}) is carried out straightforwardly.
After some algebra, one obtains
  \begin{widetext}
      \begin{eqnarray}
      \label{coh22}
   && \hspace*{-2mm}
   \langle\sigma_{\pm}(t)\rangle=\langle\sigma_{\pm}\rangle\, e^{\pm i\omega_0 t} e^{-\gamma(t)}
       \frac{\sum_m\left\{\langle 0|\Omega_m^{\dagger}\sigma_{\pm}\Omega_m|0\rangle
   e^{\beta\omega_0/2\pm i\Phi(t)}
  +\langle 1|\Omega_m^{\dagger}\sigma_{\pm}\Omega_m|1\rangle
    e^{-\beta\omega_0/2\mp i\Phi(t)}\right\}}
    {\sum_m\left\{\langle 0|\Omega_m^{\dagger}\sigma_{\pm}\Omega_m|0\rangle
   e^{\beta\omega_0/2} +\langle 1|\Omega_m^{\dagger}\sigma_{\pm}\Omega_m|1\rangle
    e^{-\beta\omega_0/2}\right\}}
 \end{eqnarray}
  \end{widetext}
 with the initial coherences
   \begin{eqnarray}
     \label{sig-NonS-init}
     &&
     \hspace*{-20pt}
     \langle \sigma^{}_{\pm}\rangle =
     \frac{1}{2\cosh\left(\beta\omega^{}_{0}/2\right)}
     \sum_m \left\{
      \langle 0| \Omega^{\dagger}_{m}\sigma^{}_{\pm}\Omega^{}_{m}|0\rangle e^{\beta\omega_0/2}\right.
      \nonumber\\[2pt]
      && \hspace*{60pt}\left.
      {}+
      \langle 1| \Omega^{\dagger}_{m}\sigma^{}_{\pm}\Omega^{}_{m}|1\rangle e^{-\beta\omega_0/2}
      \right\}.
   \end{eqnarray}
 Eq.~(\ref{coh22}) contains two relevant functions. The so-called
  \textit{generalized decoherence function\/} $\gamma(t)$ is defined as
  \begin{equation}
   \label{gamma-def}
   \gamma(t)=\int_0^{\infty}d\omega\, J(\omega)\coth(\beta\omega/2)
   \frac{1-\cos\omega t}{\omega^2},
   \end{equation}
where the spectral density $J(\omega)$, which characterizes the qubit-bath interaction, is introduced by the rule
  \begin{equation}
     \label{J(omega)}
   \sum_{k} 4|g^{}_{k}|^2\,f(\omega^{}_{k})= \int_0^{\infty} d\omega\, J(\omega) f(\omega).
   \end{equation}
In (\ref{J(omega)}), the continuum limit of the bath modes is performed.
 Usually, the spectral density function is taken in the form
\begin{equation}
\label{spectral}
J(\omega)=\lambda\,\omega_c^{1-s}\omega^s e^{-\omega/\omega_c},
\end{equation}
which is most commonly used in the theory of spin-boson
systems \cite{BPbook,luczka2013,Unruh,Leggett}.
Here $\lambda\sim|g_k|^2$ is a dimensionless coupling constant; $s$ denotes the ohmicity index, and
$\omega_c$ stands for some ``cut-off'' frequency. In other words, the constant $\lambda$ defines the interaction strength; the index $s$ determines the nature of interaction ($0<s<1$ corresponds to the sub-Ohmic case, $s=1$  to the Ohmic coupling regime, and $s>1$  to the super-Ohmic one), while $\omega_c$ defines the shape of the spectral density at large frequencies and ensure the finiteness of all the frequency moments. The spectral density function can also be modified by introducing the logarithmic perturbations of the power-law profiles \cite{giraldi2017}.

The generalized decoherence function $\gamma(t)$ is precisely  the quantity which determines
  the relaxation of the off-diagonals (\ref{coher-DM})
  due to vacuum and thermal fluctuations in the bath~\cite{PRA2012}. It follows from Eq.~(\ref{gamma-def}) that the generalized decoherence function is always positive. In the absence of initial correlations 
 \begin{equation}\label{Dgam}
\frac{d |\langle\sigma_{\pm}(t)\rangle|}{dt}=-\frac{d\gamma(t)}{dt}\left|\langle\sigma_{\pm}(t)\rangle\right|
 \end{equation}
   The time derivative of $\gamma(t)$ can change its sign, being proportional to sine function in the integrand of (\ref{gamma-def}). Usually \cite{haikka2013non,lombardo2015,giraldi2017}, the emergence of RDE is associated with the non-monotonic behaviour of $|\langle\sigma_{\pm}(t)\rangle|$ and occurs in the advanced super-Ohmic regimes. As it has been mentioned in Introduction, the number of RDE increases with $s$, and they begin to occur at shorter times.

  The other function entering Eq.~(\ref{coh22}), $\Phi(t)$, is given by
    \begin{equation}
   \label{Phi}
  \Phi(t)=\int_0^{\infty}d\omega J(\omega) \frac{\sin\omega t}{\omega^2}.
   \end{equation}
As discussed in Ref.~\cite{PRA2012}, this function accounts for the influence of the initial
 qubit-environment correlations on the dynamics of decoherence.
These correlations are inherited from the post-measurement initial
state due to the presence of the interaction term in the 
Hamiltonian (\ref{H}).

 In general, the expression (\ref{coh22}) can be rewritten
 more transparently as
    \begin{equation}
     \label{coh3}
 \langle\sigma_{\pm}(t)\rangle=\langle\sigma_{\pm}\rangle
\exp[\pm i(\omega_0 t+\chi(t)] \exp[-\gamma_{\rm tot}(t)],
    \end{equation}
where
  \begin{equation}
     \label{gamma-eff}
  \gamma_{\rm tot}(t)=\gamma(t)+ \gamma^{}_{\rm cor}(t)
    \end{equation}
is the total decoherence function including the correlation
contribution
 \begin{eqnarray}
 \label{gam_cor_NS}\nonumber
   \hspace*{-15pt}\gamma_{\rm cor}(t)=-\frac{1}{2}\ln\left\{
  1+\left(\frac{N_1^2+N_2^2}{D^2}-1\right)\sin^2\Phi(t)\right.
  \\
   \hspace*{-15pt}\left.{}+\frac{N_2}{D}\sin(2\Phi(t))\right\}.
 \end{eqnarray}
Here we have introduced
\begin{widetext}
 \begin{eqnarray}
 \label{NND}
 \nonumber
  && \hspace*{-11pt}
  N_1\!=\!\left\{e^{\beta\omega_0}\sin^4(\theta_a/2)-e^{-\beta\omega_0}\cos^4(\theta_a/2)
  \right\}\sin^2\theta_1+
    \left\{e^{\beta\omega_0}\cos^4(\theta_a/2)-e^{-\beta\omega_0}\sin^4(\theta_a/2)
  \right\}\sin^2\theta_2
 \\
\nonumber
 && \hspace*{-11pt}
 \quad\quad{} +\sinh(\beta\omega_0)\sin^2\theta_a\cos\Delta_{\phi}\sin\theta_1\sin\theta_2,
  \\[1ex]
 &&   \hspace*{-11pt}
   N_2\!=\!2\cos\theta_a\,\sin\Delta_{\phi}\, \,\sin \theta_1\, \sin \theta_2,
  \\[1ex]
  \nonumber
  && \hspace*{-11pt}
  D\!=\!\left\{\mbox{$\frac{1}{2} $}
  \sin^2\theta_a +e^{\beta\omega_0}\sin^4(\theta_a/2)+e^{-\beta\omega_0}\cos^4(\theta_a/2)
  \right\}\sin^2\theta_1 \!+\!
 \left\{\mbox{$\frac{1}{2} $}
 \sin^2\theta_a +e^{\beta\omega_0}\cos^4(\theta_a/2)+e^{-\beta\omega_0}\sin^4(\theta_a/2)
  \right\}\sin^2\theta_2
\\
  \nonumber
  && \hspace*{-10pt}
  \quad {}+\left\{\cosh(\beta\omega_0)\sin^2\theta_a
   +2\left[\sin^4(\theta_a/2)+\cos^4(\theta_a/2)\right]
       \right\}\cos\Delta_{\phi}\sin\theta_1\sin\theta_2.
  \end{eqnarray}
  \end{widetext}
 To simplify the notations, we have written $\theta^{}_{i}$,  $\phi^{}_{i}$ for
 $\theta^{}_{b_i}$, $\phi^{}_{b_i}$, and denoted $\Delta_{\phi}=\phi_1-\phi_2$.
 The expression for the phase shift $\chi(t)$ is
 \begin{equation}
 \label{chi_NS}
 \chi(t)=\arctan\left(\frac{N_1\sin\Phi(t)}{D\cos\Phi(t)+N_2\sin\Phi(t)}\right).
 \end{equation}
The initial coherences (\ref{sig-NonS-init}) can be directly evaluated as
  \begin{eqnarray}
    \label{sig0-2}
    & & \hspace*{-28pt}
   \langle\sigma_{\pm}\rangle\!=\!\frac{e^{\pm i\phi^{}_{1}}}{4\cosh(\beta\omega_0/2)}
  \nonumber\\[2pt]
  & & {}\times
   \left\{ \sin\theta^{}_{1}\!\left[ e^{\beta\omega_0/2}
     \sin^2\frac{\theta_a}{2}  +e^{-\beta\omega_0/2}\cos^2\frac{\theta_a}{2}\right] \right.
   \nonumber\\[2pt]
     & & \hspace*{-20pt}
     \left. {}\!+\!{e}^{\mp i\Delta_{\phi}}\sin\theta^{}_{2}
      \!\left[ e^{\beta\omega_0/2} \cos^2\frac{\theta_a}{2}
     \!+\! e^{-\beta\omega_0/2}  \sin^2\frac{\theta_a}{2}\right] \right\}.
  \end{eqnarray}
Equations (\ref{coh3})\,--\,(\ref{sig0-2}) determine  the time-dependent coherences
 $\langle\sigma^{}_{\pm}(t)\rangle $ or, what is the same, the off-diagonals of the
  qubit density matrix [see Eq.~(\ref{coher-DM})].

In a similar way, one can also calculate the average values $\langle\sigma^{}_{3}\rangle$ for 
the general non-selective measurement scheme by using a formula similar to (\ref{sig-NonS-init}).
 A straightforward algebra gives
   \begin{eqnarray}
   \label{sigZ}
   \nonumber & & \hspace*{-30pt}
   \langle\sigma_{3}\rangle= \frac{1}{2\cosh(\beta\omega_0/2)}
   \\[2pt]
   \nonumber & & \hspace*{-20pt}
    {}\times\left\{\cos\theta_1\left[ e^{\beta\omega_0/2}\sin^2\frac{\theta_a}{2}
   + e^{-\beta\omega_0/2}\cos^2\frac{\theta_a}{2}\right]\right.
   \\[2pt]
   & & \hspace*{-10pt} \left.
   {}+\cos\theta_2 \left[e^{\beta\omega_0/2}\cos^2\frac{\theta_a}{2}
    + e^{-\beta\omega_0/2}\sin^2\frac{\theta_a}{2}\right]\right\}.
 \end{eqnarray}
 
 Two important conclusions can be made, when inspecting Eqs.~(\ref{gamma-def}) and (\ref{gam_cor_NS}). While the combined action of the vacuum and thermal fluctuations is of the first order in the coupling constant $\lambda$ (or, what is the same, of the second order in the spin-boson interaction strength $|g_k|^2$), the correlation contribution $\gamma_{\rm cor}(t)$ contains all orders in $\lambda$, being a non-linear function of $\Phi(t)$. The same is true for the phase shift (\ref{chi_NS}), which defines the  renormalization of the qubit energy due to the initial correlations with the thermal bath \cite{PRA2012,myPRA2015}. 

On the other hand, it is seen from (\ref{gam_cor_NS}) that the correlation contribution $\gamma_{\rm cor}(t)$ to the generalized decoherence function (\ref{gamma-eff}) oscillates with time. Depending on the initial qubit preparation (expressed via the coefficients $N_1$, $N_2$ and $D$), it can change its sign at some particular times, if the coupling strength $\lambda$ is large enough to provide the proper values of the sine functions. It means that the periods of decoherence and recoherence can change each other \cite{giraldi2021,giraldi2022short}. The first scenario occurs when the total decoherence function is positive, $\gamma_{\rm tot}(t)>0$. In the opposite case, if $\gamma_{\rm tot}(t)<0$, the coherence enhancement takes place \cite{myPRA2015,overcomplete,giraldi2022short} at small times, and one can speak about calculation of the coherence increment rather than decrement.
At larger times, the joint contribution of the vacuum and thermal fluctuations expressed by (\ref{gamma-def}) definitely exceeds the effect of $\gamma_{\rm cor}(t)$, making the total decoherence function (\ref{gamma-eff}) positive, leading to either complete or partial system decoherence  \footnote{It was shown in Ref.~\cite{PRA2012} that the value given by Eq.~(\ref{gamma-def}) is an increasing function of time at the ohmicity indexes $0<s<2$, whereas in the advanced super-Ohmic regime $s\ge 2$ it saturates at large times.}.

Rather cumbersome expressions (\ref{gam_cor_NS})-(\ref{NND}) can be simplified considerably using the following reasoning.
Note, that the correlation contribution in Eq.~(\ref{coh22}) can be rewritten
as \begin{equation}\label{cor2} \exp[\pm
i\chi(t)-\gamma_{\rm cor}(t)]=\cos\Phi(t)\pm i {\cal
	A}\sin\Phi(t),\end{equation}
where 
\bea \label{defC} \nonumber {\cal
	A}\!\!=\!\frac{\sum\limits_{m}\!\{\langle
	0|\Omega_m^{\dagger}\sigma_{\pm}\Omega_m |0\rangle
	e^{\beta\omega_0/2}\!-\!\langle
	1|\Omega_m^{\dagger}\sigma_{\pm}\Omega_m |1\rangle
	e^{-\beta\omega_0/2}\}}{\sum\limits_{m}\!\{\langle
	0|\Omega_m^{\dagger}\sigma_{\pm}\Omega_m |0\rangle
	e^{\beta\omega_0/2}\!+\!\langle
	1|\Omega_m^{\dagger}\sigma_{\pm}\Omega_m |1\rangle
	e^{-\beta\omega_0/2}\}}.\\ \eea 
It is seen from Eq.~(\ref{defC})
that the equality \bea\label{dual}\nonumber\sum\limits_{m}\{
\langle 0|\Omega_m^{\dagger}\sigma_{\pm}\Omega_m 0|\rangle
+\langle 1|\Omega_m^{\dagger}\sigma_{\pm}\Omega_m 1|\rangle\} \\=
\mbox{Tr}_S\left(\sigma_{\pm}\sum_{m}\Omega_m\Omega_m^{\dagger}\right)=0
\eea gives the value ${\cal A}=\coth(\beta\omega_0/2)$. The last
row in Eq.~(\ref{dual}) means that the Gram operator has to be dia\-go\-nal. 
At such a condition, Eq.~(\ref{gam_cor_NS})
is manipulated to a much simpler  form
\begin{equation}
\label{gam_cor_NS1}
\gamma_{\rm cor}(t)=
-\frac{1}{2}\ln\left[1+\frac{\sin^2\Phi(t)}{\sinh^2(\beta\omega_0/2)} \right].
\end{equation}
It can be shown to correspond to the case $N^{}_{2}=0$, see Eqs.~(\ref{gam_cor_NS})-(\ref{NND}). 

Since the values of (\ref{gam_cor_NS1}) are non-positive at any time, this condition 
ensures the coherence enhancement at the initial stage of the system evolution if $\gamma_{\rm tot}(t)<0$,
until the vacuum and
thermal fluctuations of the bath terminate the growth of
$\langle\sigma_{\pm}(t)\rangle$ and give rise to the decoherence
processes with  $\gamma_{\rm tot}(t)>0$.
If condition (\ref{dual}) is not fulfilled, the form of ${\cal A}$ is much more complicated, yielding the generic expressions (\ref{gam_cor_NS})-(\ref{NND}).
In such a case, the domains
of eventual coherence enhancement at some times can be
interchanged with the regions of the intensified decoherence at
other times \cite{giraldi2021,giraldi2022short}, depending on the values of the measurement state
vectors, the qubit-bath interaction strength, the ohmicity index and temperature.

To conclude this Section, let us make two essential remarks. First of all, unlike the commonly accepted approach \cite{lombardo2015}, we will associate the RDE with termination of the negative branch of $\gamma_{\rm tot}(t)$, rather than with a non-monotonic behaviour of $|\langle\sigma_{\pm}(t)\rangle|$. In other words, we are interested in the periods of time, when the running value of $|\langle\sigma_{\pm}(t)\rangle|$ exceeds its initial value at $t=0$; we call this effect  the recoherence (or coherence enhancement). More precisely, such a definition is suitable for treating the recoherence phenomenon as \textit{the relative coherence enhancement} at short times.

Secondly, it is seen that under the condition ({\ref{dual}}), the recoherence increment (\ref{gam_cor_NS1}) does not depend on the initial state of the qubit. It differs from both the general case of non-selective measurements (\ref{gam_cor_NS}) and the initial preparation of the system by the selective measurement (see Eq.~(31) in Ref.~\cite{PRA2012}), when the qubit-bath initial correlation opens an additional channel of the system decoherence, which depends on the mean value $\langle\sigma_3\rangle$. Since at the condition (\ref{dual}) both components of the function (\ref{gamma-eff}) depend only on the state of the thermal bath, an interesting problem arises: to study how the parameters of the qubit environment (the interaction strength, the ohmicity, and the bath temperature) influence the coherence enhancement at the initial state of the system evolution. An answer to this question is the topic of the Section \ref{secIV}.

\section{An influence of the thermal bath characteristics on the qubit coherence enhancement \label{secIV}}

Before presenting results for the recoherence times and recoherence increments, we would like to explain our motivation to focus on the non-selective measurements with the diagonal Gram operator (including the case $G=I$, see Eq.~(\ref{Phi-norm-cond})). 
Though the general case (\ref{gam_cor_NS}) exhibits much richer RDE dynamics, it is the choice (\ref{gam_cor_NS1}) that allows one to exclude the influence of the spin state on the time behaviour of the open quantum system at the initial stage of its evolution and to study solely the action of environment. In the case under study, the spin state defines only the initial values of the coherence, which are considerably simplified at the measurement scheme with $G=\mbox{diag}$ (see, e.~g., Eqs.~(61) and (67) in Ref.~\cite{myPRA2015}). 

However, there are also some other cases, when it becomes crucial to ``separate'' the initial spin state from that of the bath as much as possible. For instance, at the study of time behaviour of the quantum concurrence of two non-interacting spins within the identical dephasing model \cite{entangl2015},  the pure non-entangled initial spin state was chosen as $\varrho_S(0)=|+,+\rangle\langle +,+|$, where $|+\rangle$ denotes the eigenstate of the 1st Pauli matrix, $\sigma_1|+\rangle=|+\rangle$. Under such a choice, all elements of the initial spin density matrix $\varrho_S(0)$ are equal to 1, and the matrix is multiplied by~$1/4$ to fulfil the normalization condition $\mbox{Tr}_S \varrho_S(0)=1$. Since the eigenvalue calculation, arising in such a problem \cite{Wooters1,Wooters2}, is very sensitive to the coefficients at the elements of the reduced density matrix, and it is essential for the initial two-qubit state not to be part of a decoherence-free subspace, the above mentioned choice looks optimal if one focuses solely on the study of the bath influence on the qubits entanglement.

The interplay between recoherence and decoherence in the general case (\ref{gam_cor_NS}) was thoroughly studied in Ref.~\cite{giraldi2021}: the total number of RDE depending on the qubit and bath parameters was calculated, and the numerical and analytic results for the recoherence times at the Ohmic coupling were obtained. 
However, an analytical study of RDE is limited to essentially short times \cite{giraldi2022short}, when the recoherence times can be expressed via the frequency moments $\langle\omega^n\rangle\equiv\int_0^{\infty} \omega^n J(\omega) d\omega$ with $n=1,2$. But even then, the large number of possible initial qubit states does not allow one to systemize the obtained results with respect to the influence of the thermal bath on the coherence dynamics.

To begin with the study of the recoherence processes, let us remind that the joint vacuum and thermal contribution (\ref{gamma-def}) to the total decoherence function (\ref{gamma-eff}) is linear in the coupling strength,  while the times series expansion of its correlation counterpart (\ref{gam_cor_NS1}) starts from the quadratic term and contains all even power in $\lambda$ terms. Thus the coherence enhancement is a threshold phenomenon: at small qubit-bath interaction, the linear term associated with (\ref{gamma-def}) prevails, and we have pure decoherence at all times. At larger coupling strengths, $\gamma_{\rm cor}(t)$  dominates, ensuring the coherence enhancement at small times. The threshold value $\lambda_{\rm min}$, which separates the pure decoherence behaviour at $\lambda<\lambda_{\rm min}$ from the coherence enhancement at $\lambda>\lambda_{\rm min}$ can be find analytically by equating to zero the second derivative of the function (\ref{gamma-eff}) with respect to time.
After some algebra, one finds
\bea\label{lam-min}
\nonumber
\lambda_{\rm min}(s,T)=\frac{(T/\omega_c)^{s+1}\Gamma(s+1)\sinh^2(\omega_0/2 T)}{(s-1)^2\Gamma(s-1)^2}\\
\times \left\{\zeta(s+1,1+T/\omega_c)+\zeta(s+1,T/\omega_c)
\right\},
\eea
where $\zeta(s,a)$ denotes the generalized Riemann zeta function.

\begin{figure}[htb]
	\centerline{\includegraphics[height=0.28\textheight,angle=270]{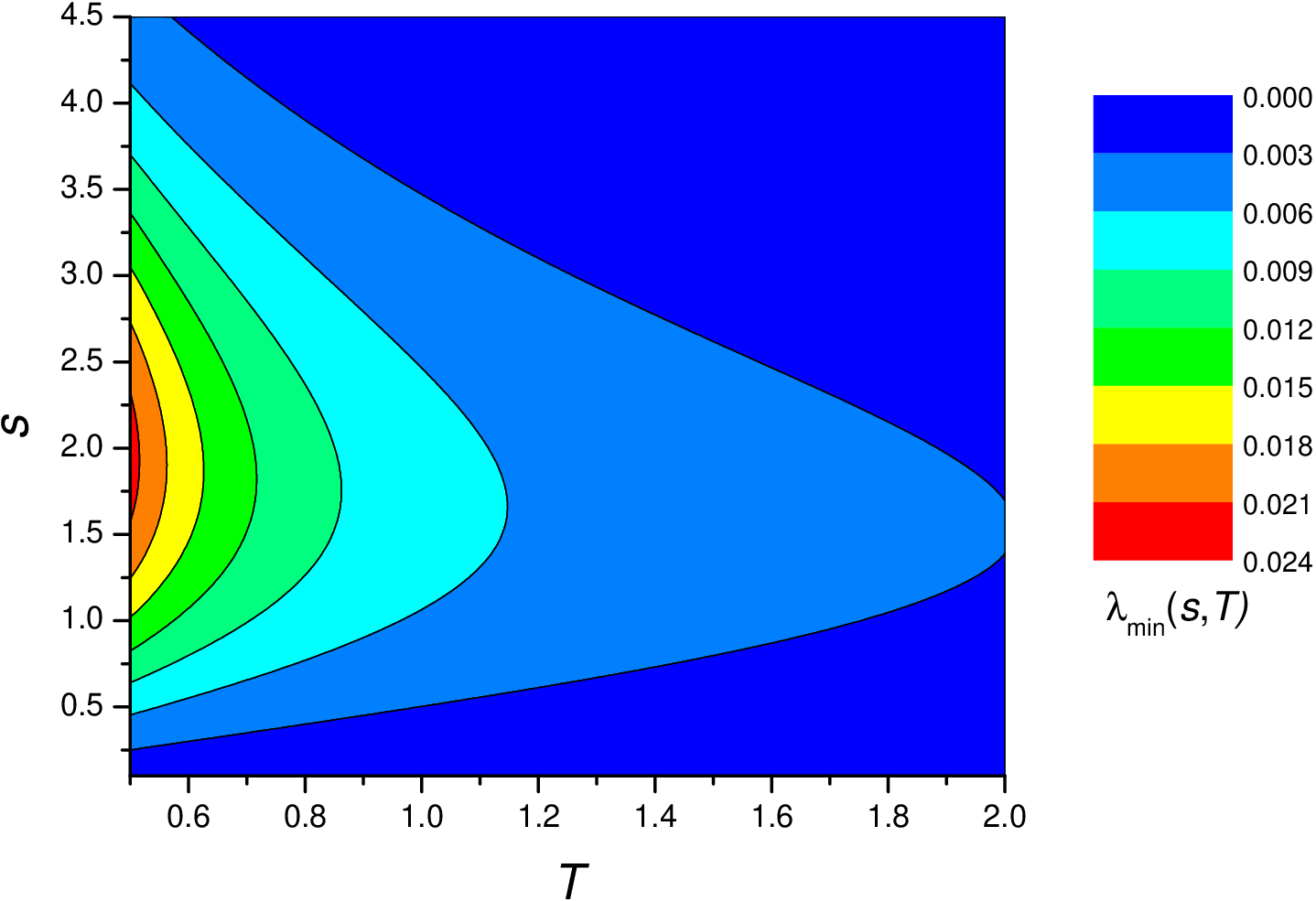}}
	\caption{The contour plot of the critical coupling constant $\lambda_{\rm min}(s,T)$ as the function of temperature $T$ and ohmicity parameter $s$. The moderate temperature regime is displayed.}
	\label{lammin}
\end{figure}

In Fig.~\ref{lammin}, the contour plot describing the ($s-T$) dependence of the critical coupling constant $\lambda_{\rm min}$ is displayed. 
Hereafter, all the parameters are presented in the dimensionless form with respect to the cut-off frequency $\omega_c$, namely $\omega_0\rightarrow\omega_0/\omega_c$, $T\rightarrow T/\omega_c$, and $t\rightarrow\omega_c t$. Besides, we take $\omega_c=1$ and $\omega_0=0.1$ in all our calculations.

One can see that the maximum values of $\lambda_{\rm min}$ shift with temperature from the advanced super-ohmic regime with $s>2$ to  about $s=1.5$. It is obvious that high temperatures are more favourable for smaller values of $\lambda_{\rm min}$, while at low $T$ the recoherence occurs at the strong coupling only. It is quite expected, since according to Eq.~(\ref{gam_cor_NS1}) the correlation contribution $\gamma_{\rm cor}(t)$ decreases at low temperatures, and there is no other option but to increase the qubit-bath interaction to maintain the eventual recoherence.

Quite remarkably (although coincidentally) the locations of maxima resemble the distribution of $s$ and $\beta$, which corresponds to the maxima of the qubits concurrence in the dephasing model, see Fig.~7 of Ref.~\cite{entangl2015}. These maximum values were found to occur in the narrow domain  $s\sim 3$, being slightly dependent on temperature. Therefore, the advanced super-Ohmic regime appears to be the most suppressive for the coherence enhancement at small times and the most favourable for the qubits states entanglement.

Now let us focus our attention on the time dependence on the total decoherence function (\ref{gamma-eff}). As has been already said, the negative values of $\gamma_{\rm tot}(t)$ correspond to the coherence enhancement and define the coherence increment. On the contrary, the case $\gamma_{\rm tot}(t)>0$ is related to the decoherence and defines its decrement.

  \begin{figure}[htb]
 	\centerline{\includegraphics[height=0.22\textheight,angle=270]{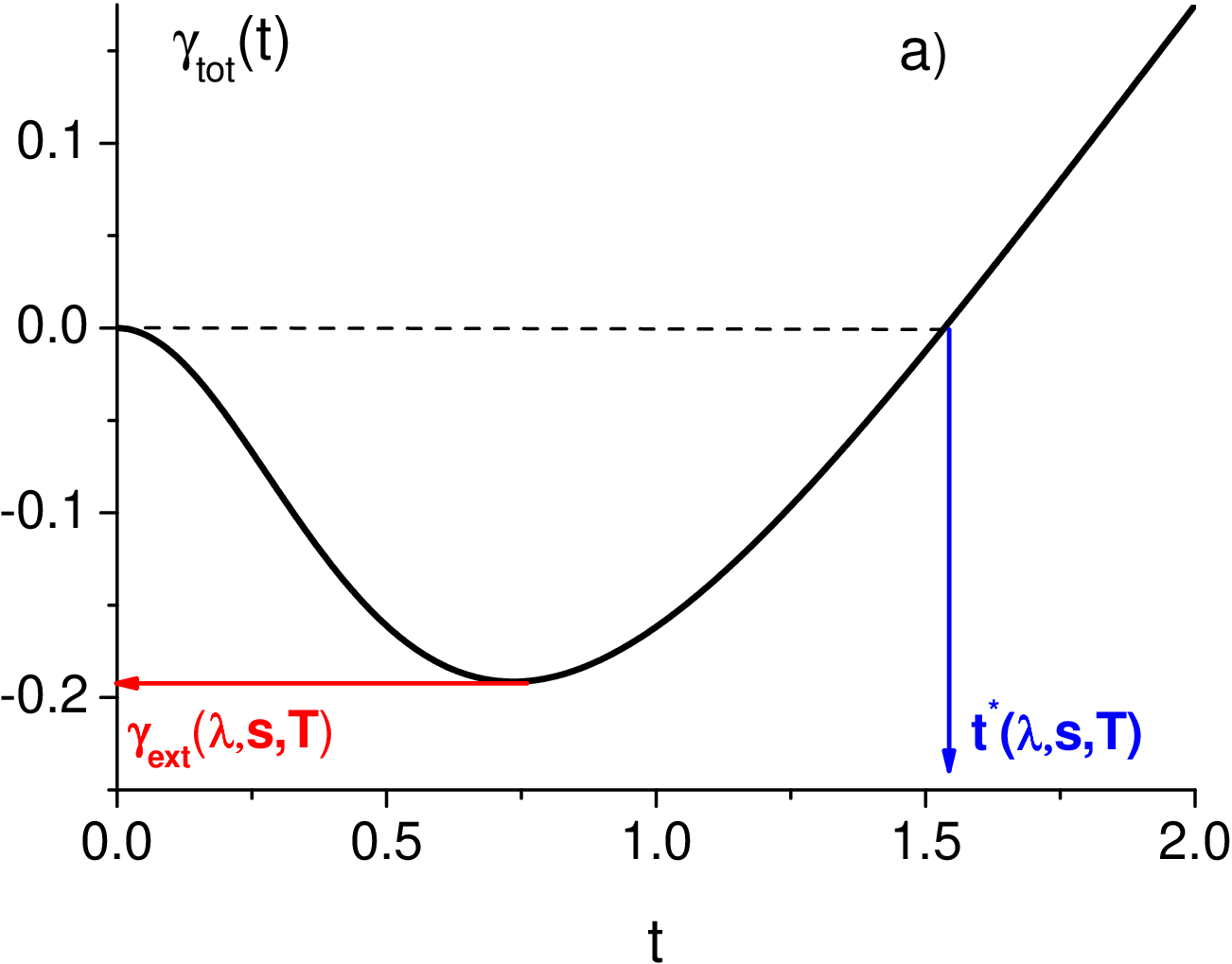}}
 	\includegraphics[height=0.22\textheight,angle=270]{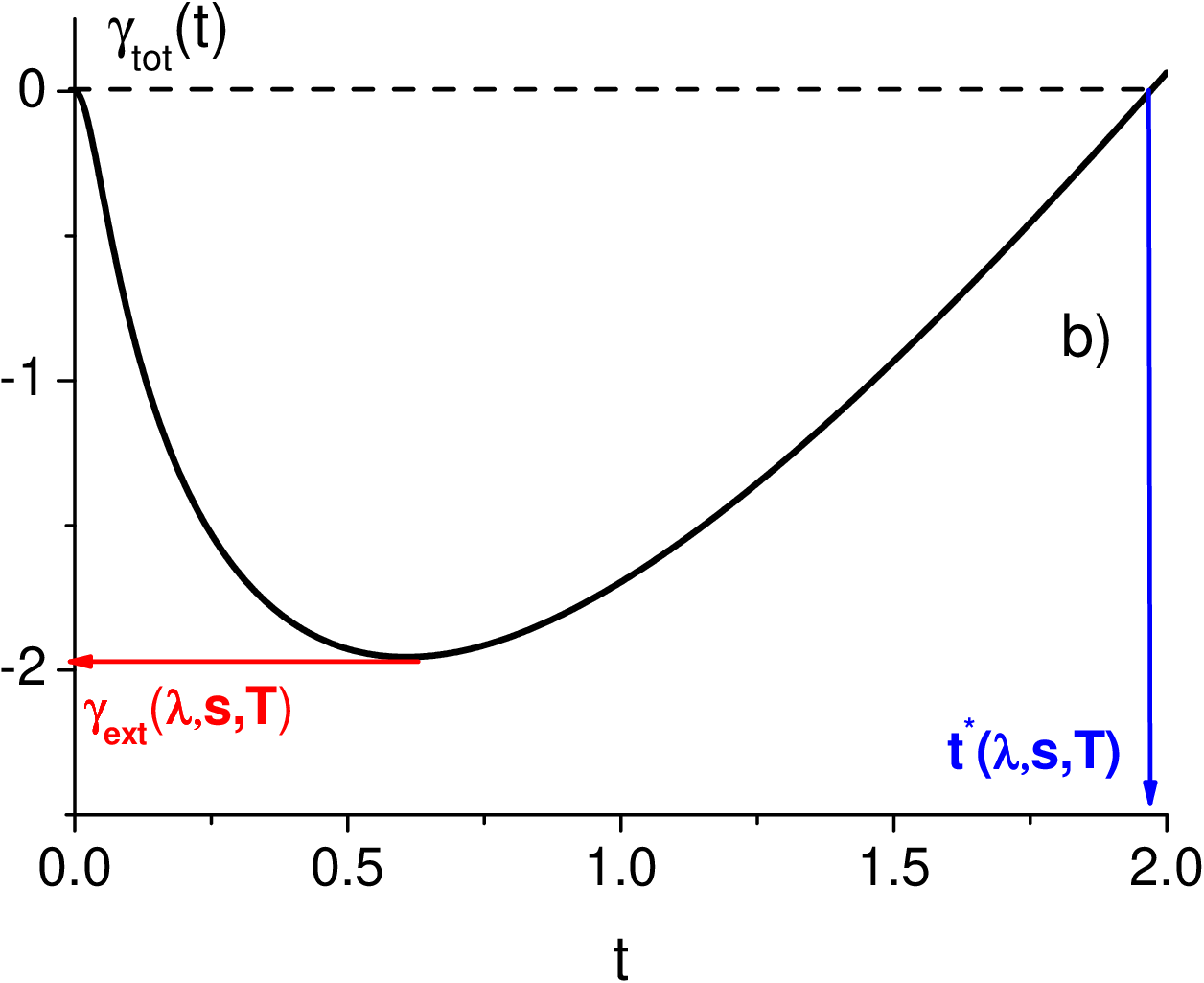}
 	\hspace*{2mm}\includegraphics[height=0.22\textheight,angle=270]{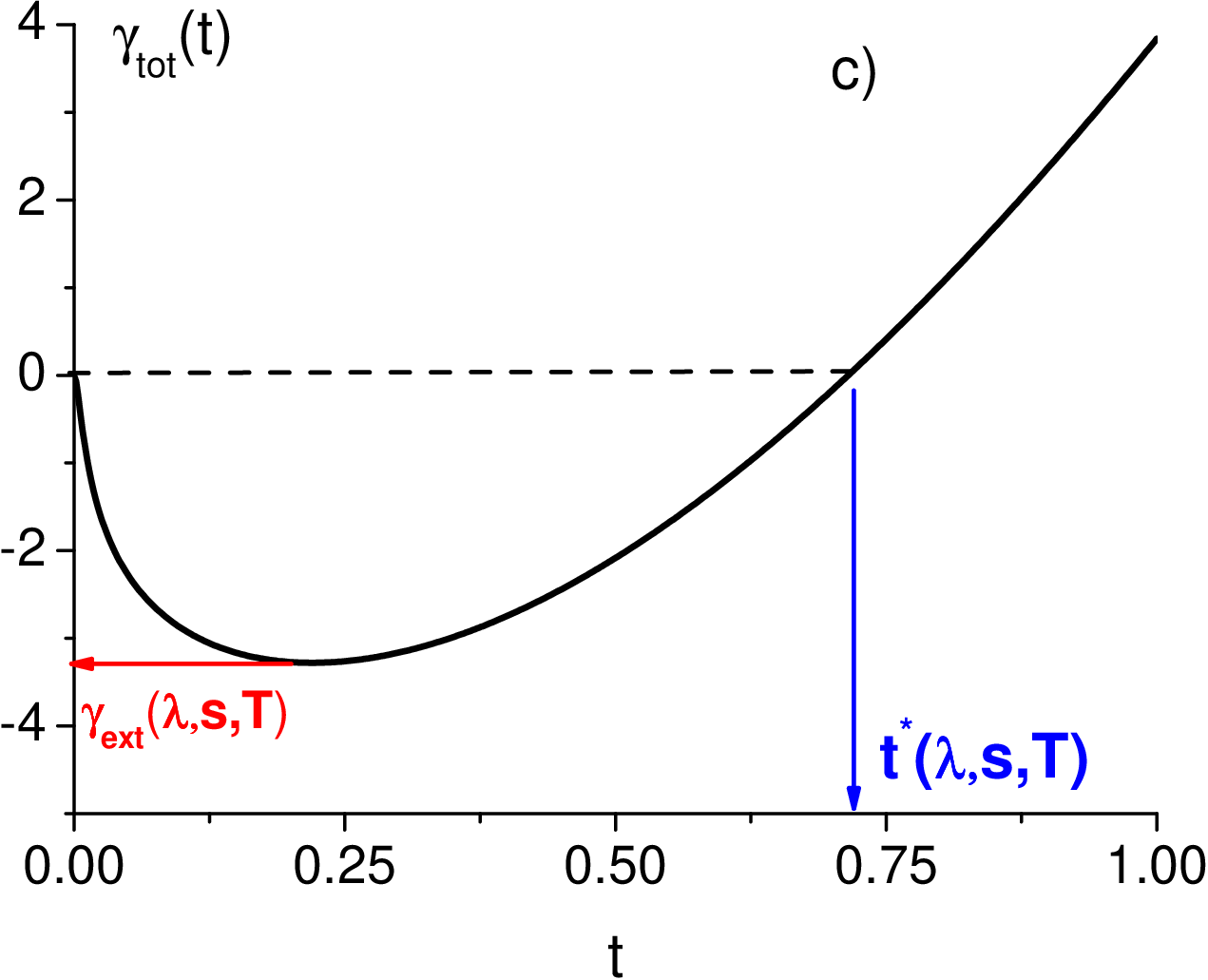}
  	\caption{Time dependence of $\gamma_{\rm tot}(t)$ at the Ohmic regime $s=1$ and intermediate coupling $\lambda=1$ at three different temperatures: a) $T=0.1$, b) $T=1$ and c) $T=10$.}
 	\label{3gamma}
 \end{figure}

In Fig.~\ref{3gamma}, we show the time behaviour of $\gamma_{\rm tot}(t)$ at three different temperatures in the Ohmic regime and intermediate coupling. Since the coupling strength $\lambda=1$ exceeds the critical value $\lambda_{\rm min}$, one observes the initial recoherence, lasting until the curve $\gamma_{\rm tot}(t)$ intersects the abscissa axis. We marked this coordinate, indicated  by the blue arrow, as $t^*(\lambda,s,T)$ and defined this value as the \textit{recoherence time}. 
Correspondingly, the minimal value of  (\ref{gamma-eff}), indicated by the red arrow, is labelled as $\gamma_{\rm extr}$. 
This value is a gradually decreasing function of temperature, as it directly follows from Eq.~(\ref{gam_cor_NS1}). It defines the maximum coherence increment $\Gamma_{\rm max}=|\gamma_{\rm extr}|$, also gradually increasing with $T$.
%

It has to be emphasized that within the usual definition of the RDE \cite{lombardo2015,giraldi2017}, the first recoherence/decoherence event begins at the moment, when 
the time derivative of $\gamma_{\rm tot}(t)$ changes its sign. 
To comply with this widely accepted definition, we would have to redefine the recoherence time as the time indicated by the red arrow rather than blue one in Fig.~\ref{3gamma}. 
With this redefinition, the general results for $t^*$ shown later in the top panels of Figs.~\ref{tg-s} and \ref{tg-T} would not change essentially, and all the basic qualitative trends would remain the same.
%
 \begin{figure}[htb]
	\centerline{\includegraphics[height=0.25\textheight,angle=270]{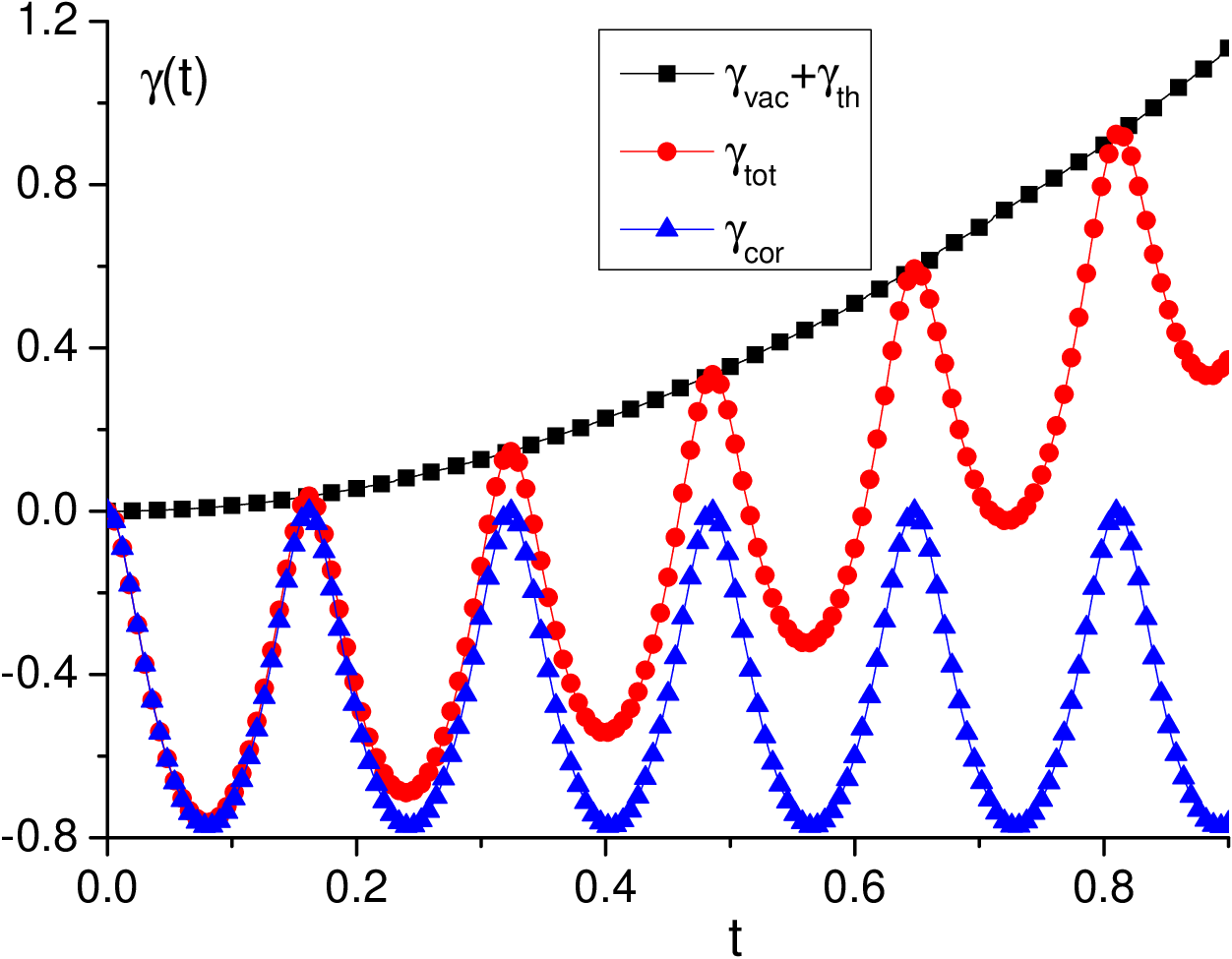}}
		\centerline{\includegraphics[height=0.25\textheight,angle=270]{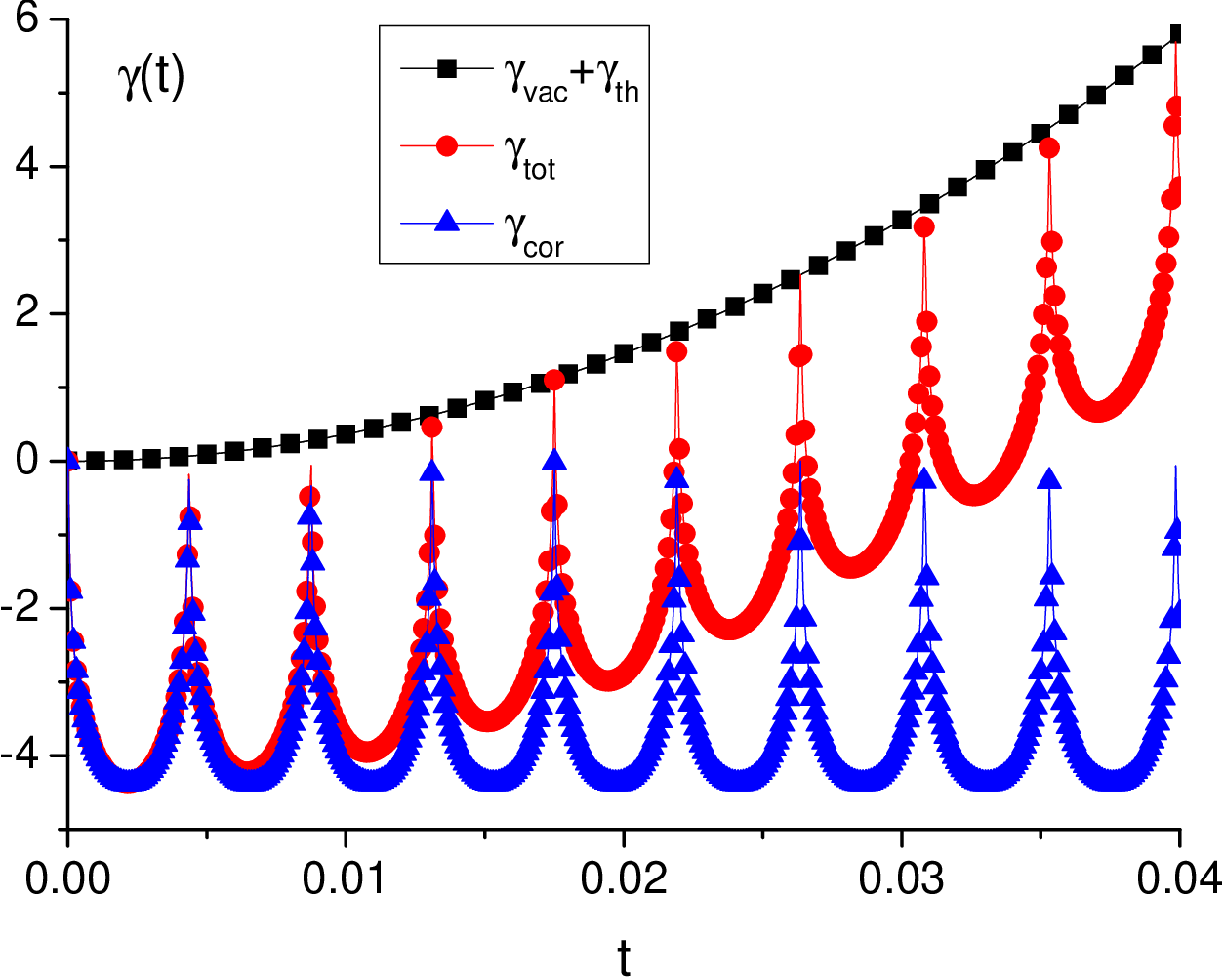}}
	\caption{Time dependence of different contributions to the generalized decoherence function, calculated at $\lambda=1$, $s=0.05$, $T=0.1$ (top panel), and $\lambda=1$, $s=7$, $T=4$ (bottom panel). Red circles correspond to the total decoherence function (\ref{gamma-eff}), black squares to the sum of the vacuum and thermal contributions (\ref{gamma-def}), and blue triangles to the correlation contribution (\ref{gam_cor_NS1}). 
	}
	\label{2gamma-oscil}
\end{figure}

At the special choice of the bath parameters, there appears a sequence of the RDE at short times, as it is clearly seen in Fig.~\ref{2gamma-oscil}. The oscillatory behaviour of $\gamma_{\rm cor}(t)$ is strongly enhanced both in the advanced sub-Ohmic and super-Ohmic regimes, since the function (\ref{Phi}) (the argument of the sine in Eq.~(\ref{gam_cor_NS1})) diverges as $s^{-1}$ at small ohmicity indexes and grows as $(s-2)!$ at large integer $s$. It can be easily verified by expanding the $\Gamma$-functions entering the expressions for $\Phi(t)$, see, e.~g., Ref.~\cite{PRA2012}. 

The vacuum and thermal fluctuations of the bath yield the gradually increasing generalized decoherence function (\ref{gamma-def}), which forms the envelope curve touching $\gamma_{\rm tot}(t)$ at its maxima points. These maxima are smooth (top panel) or sharp (bottom panel) depending on the values of $s$ and the bath temperature. However, in both cases a strongly oscillating correlation contribution $\gamma_{\rm cor}(t)$ to the total decoherence function forms a sequence of the RDE. In our definition, the recoherence time $t^*$ corresponds to the first intersection of the abscissa axis by the red line.

One can also introduce a notion of the \textit{total recoherence time}. Let us consider the time intervals $\tau_i=[t_{i-1},t_i]$ formed by two successive instants $t_{i-1}$ and $t_i$, between which $\gamma_{\rm tot}(\tau_i)<0$.
The function $\gamma_{\rm tot}(t)$ can be shown to have a negative branch only if
\be\label{oscil}
\gamma(t)\le \ln[\coth(\beta\omega_0)/2].
\ee
 The expression in the r.h.s. of the inequality (\ref{oscil}) corresponds to the maximum value of $|\gamma_{\rm cor}(t)|$, as it can be easily derived from Eq.~(\ref{gam_cor_NS1}).
 In other words, the number of recoherence events equals the maximum value $i_{\rm max}$ of the integer $i$, at which Eq.~(\ref{oscil}) still holds true. 
 
 Now it is natural to introduce the total recoherence time as $t^*_{\rm tot}=\sum_{i=1}^{i_{\rm max}}\tau_i$.
 It is seen from the bottom panel of Fig. \ref{2gamma-oscil} that in the advanced super-Ohmic regime at high temperature, the total recoherence time is approximately equal to $(i_{\rm max}-1)$ times the period of $\gamma_{\rm cor}(t)$, since the total decoherence function $\gamma_{\rm tot}(t)$ changes sharply at its upper part. In the advanced sub-Ohmic regime and at low temperatures (top panel in Fig.~\ref{2gamma-oscil}), the behaviour of $\gamma_{\rm cor}(t)$ is rather smooth. The total recoherence time has to be calculated numerically as the sum of all $\tau_i$.

\begin{figure*}[htb]
	\centerline{\includegraphics[height=0.24\textheight,angle=270]{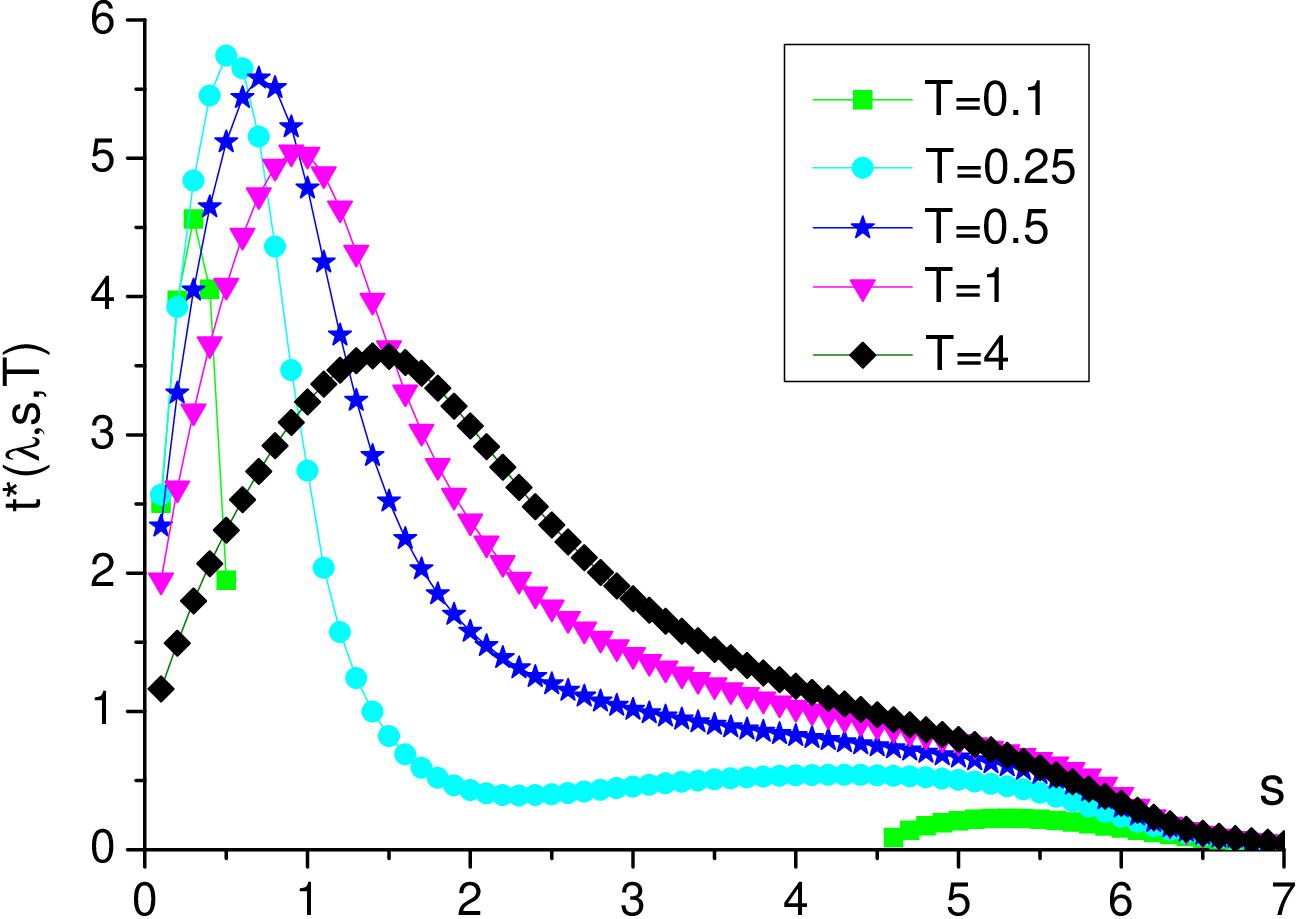}
		\includegraphics[height=0.24\textheight,angle=270]{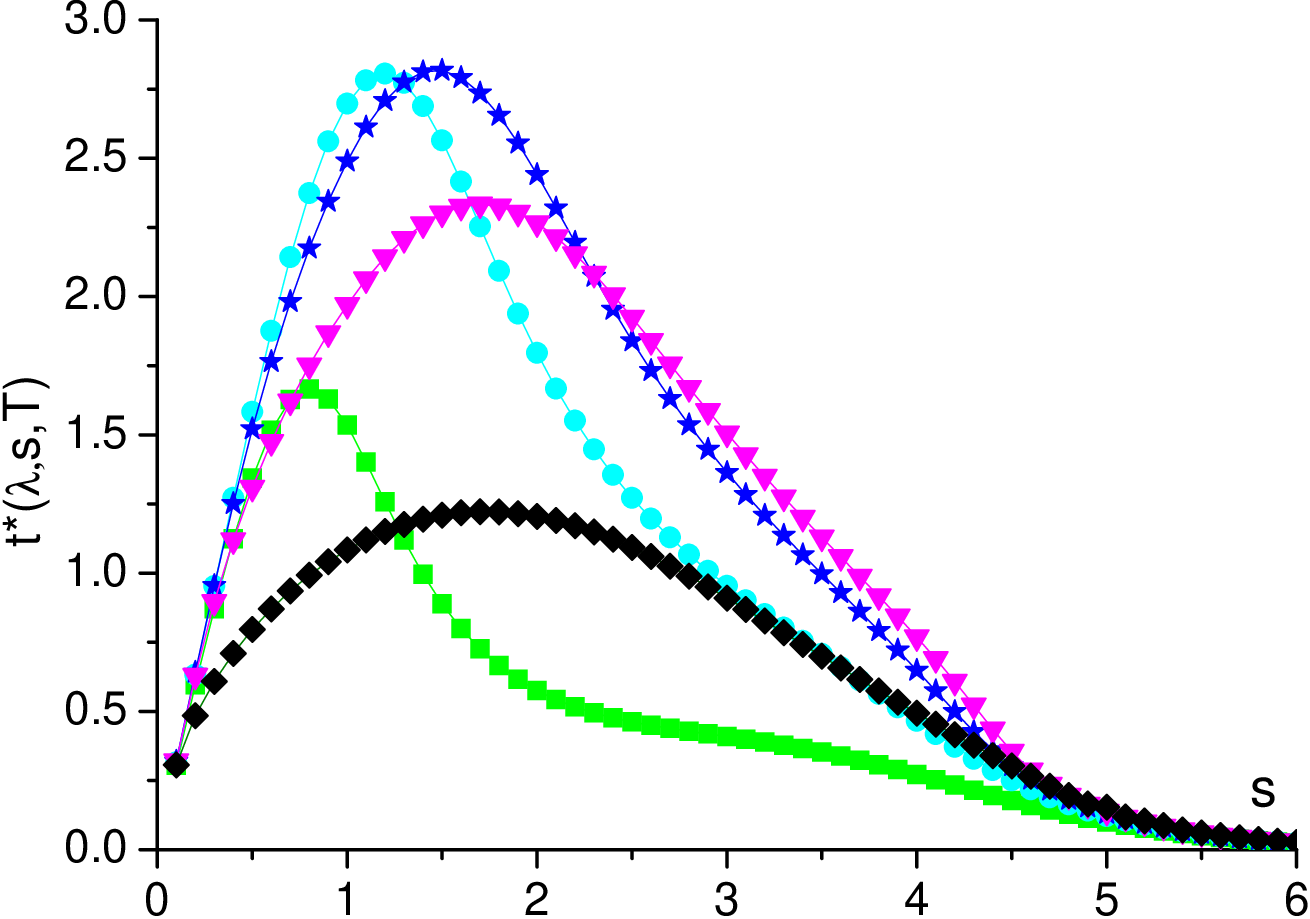}
		\includegraphics[height=0.24\textheight,angle=270]{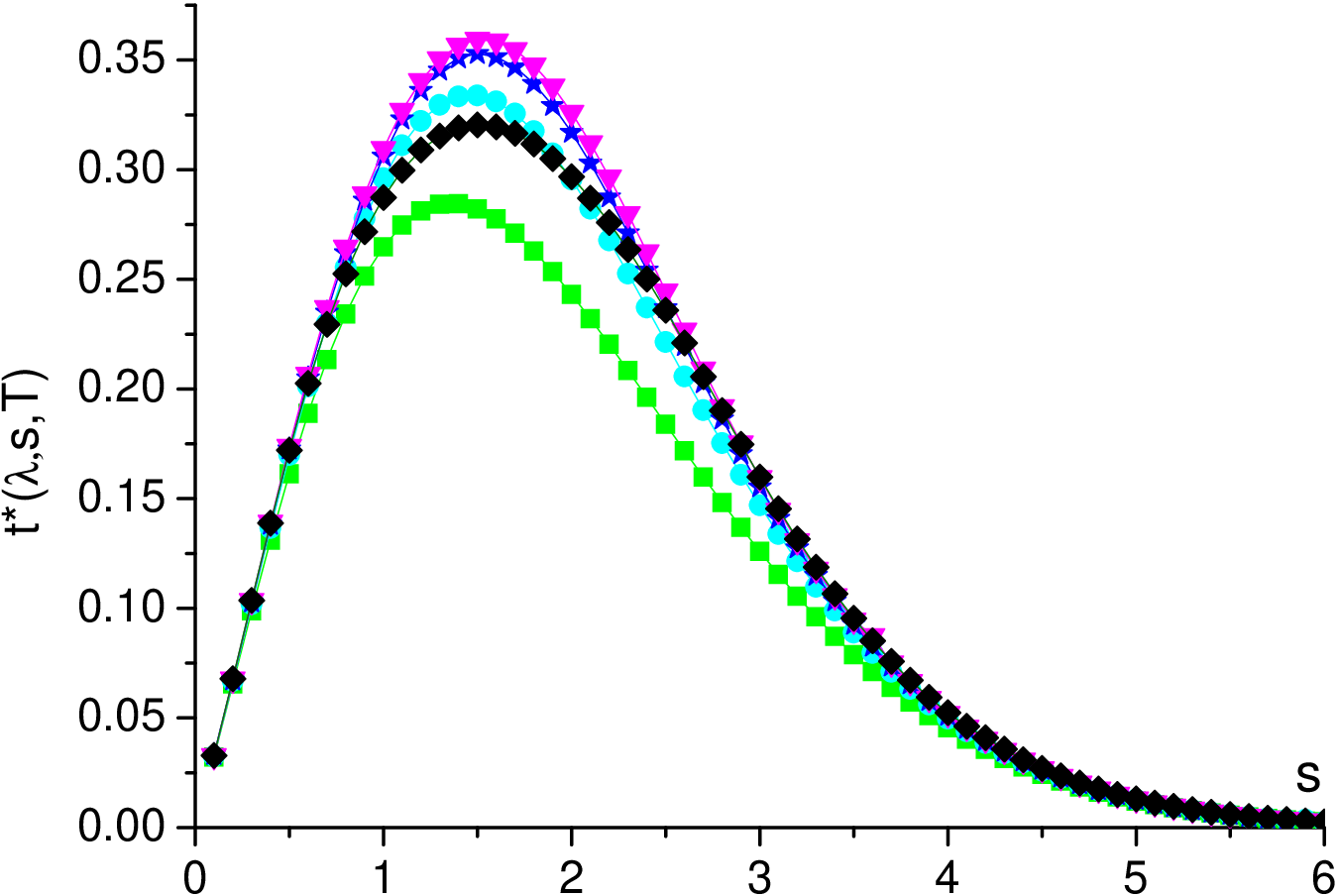}}
	\vspace*{3mm}
		\centerline{	
			\hspace*{3mm}
	\includegraphics[height=0.24\textheight,angle=270]{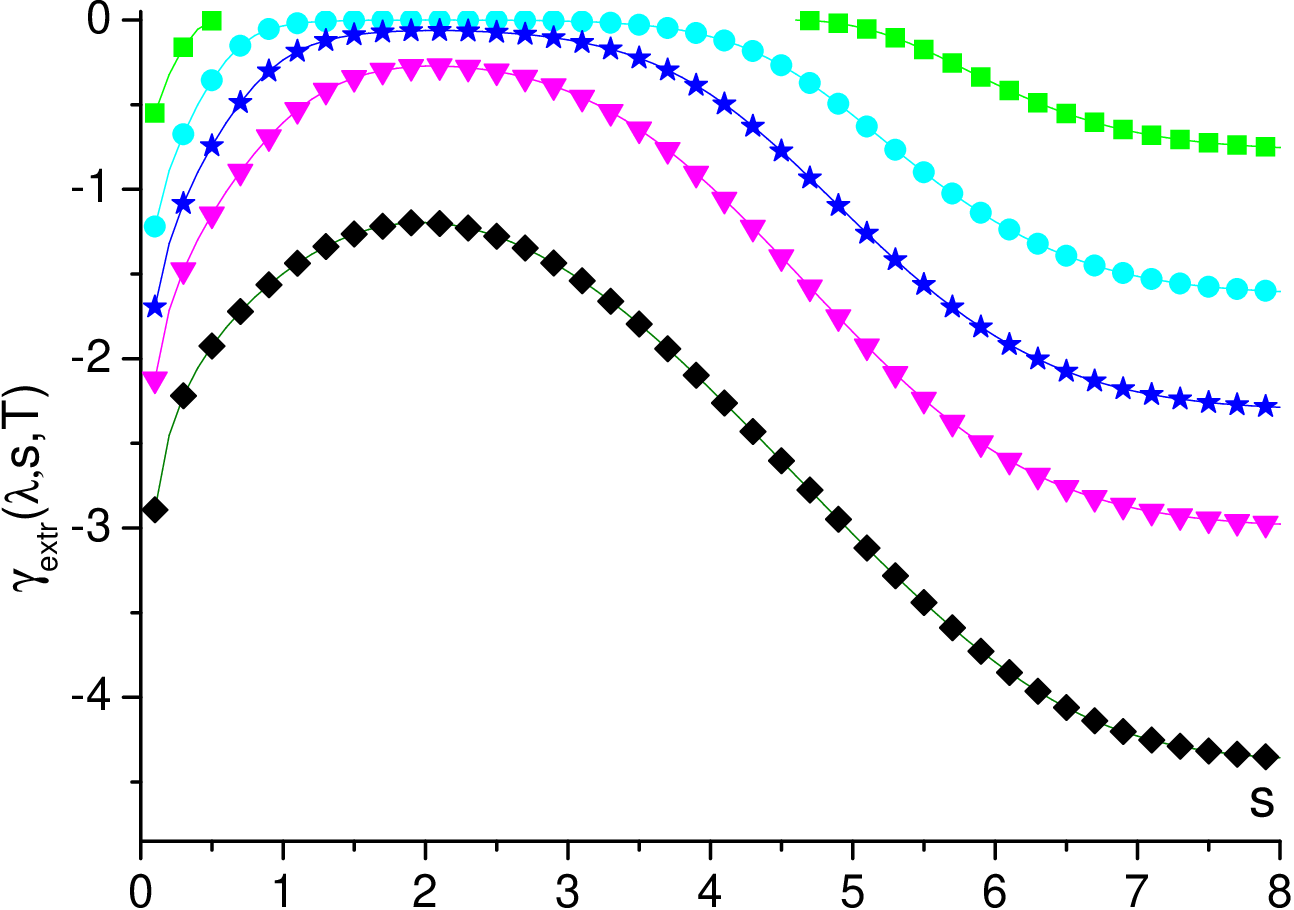}
\includegraphics[height=0.24\textheight,angle=270]{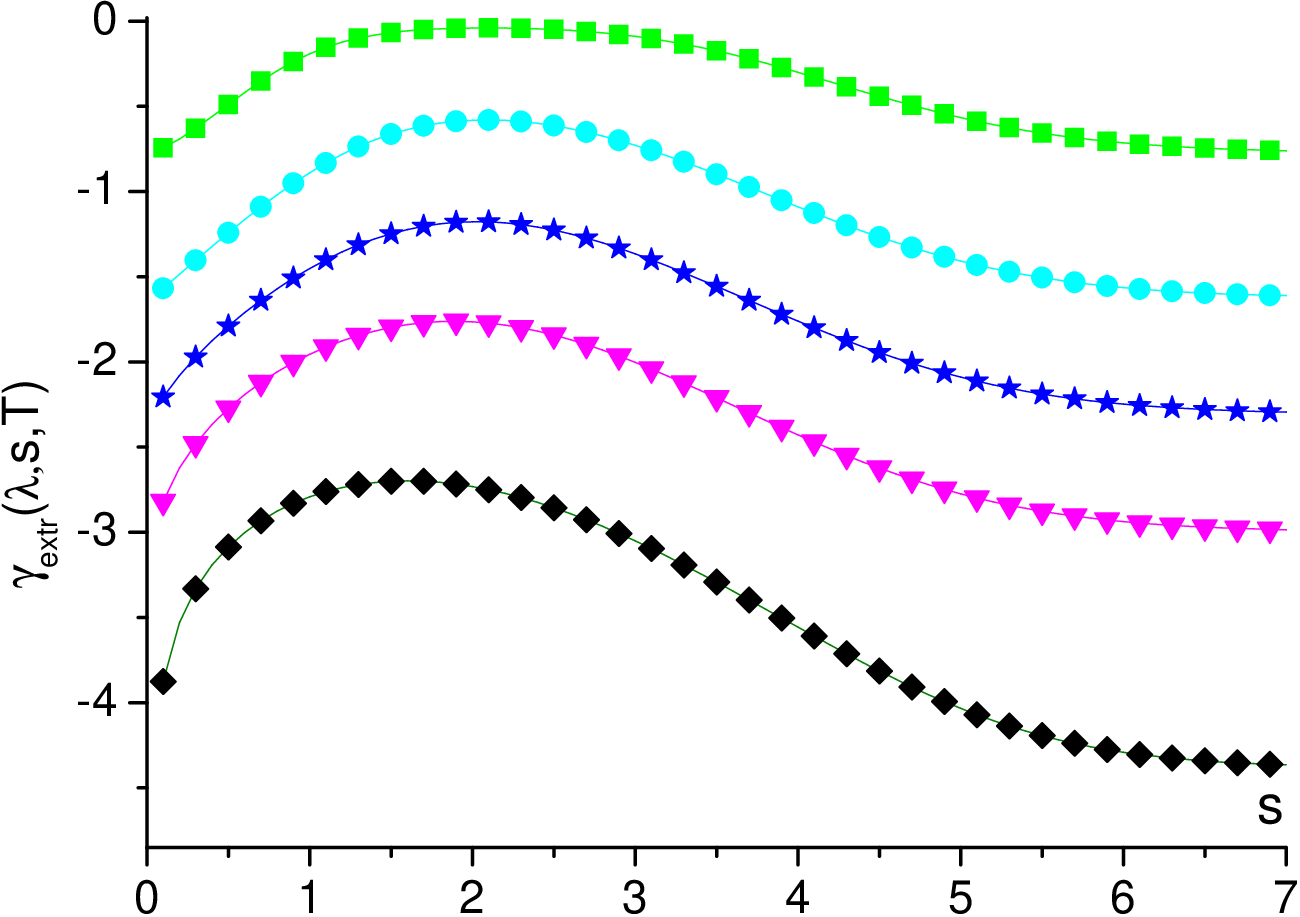}
	\includegraphics[height=0.25\textheight,angle=270]{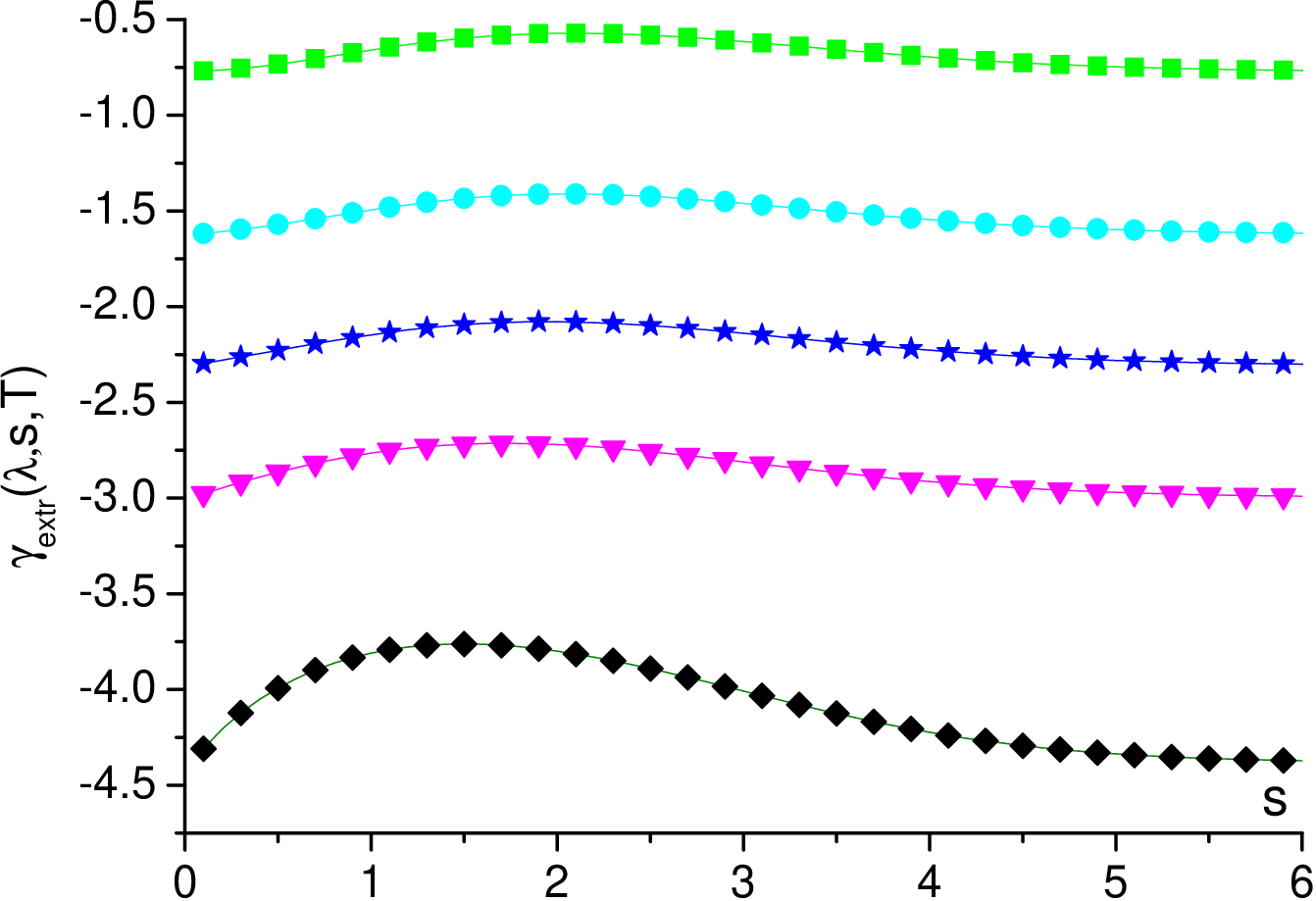}
}
	\caption{Top row: the recoherence times $t^*(\lambda,s,T)$ in the low (left panel, $\lambda=0.1$), moderate (central panel, $\lambda=1$), and high (right panel,  $\lambda=10$) coupling regimes at different temperatures $T$ as functions of the ohmicity index $s$. Bottom row: the same for the minimum values $\gamma_{\rm extr}(\lambda, s,T)$ of the total decoherence function $\gamma_{\rm tot}(t)$.
	}
	\label{tg-s}
\end{figure*}
%
\begin{figure*}[htb]
	\centerline{\includegraphics[height=0.24\textheight,angle=270]{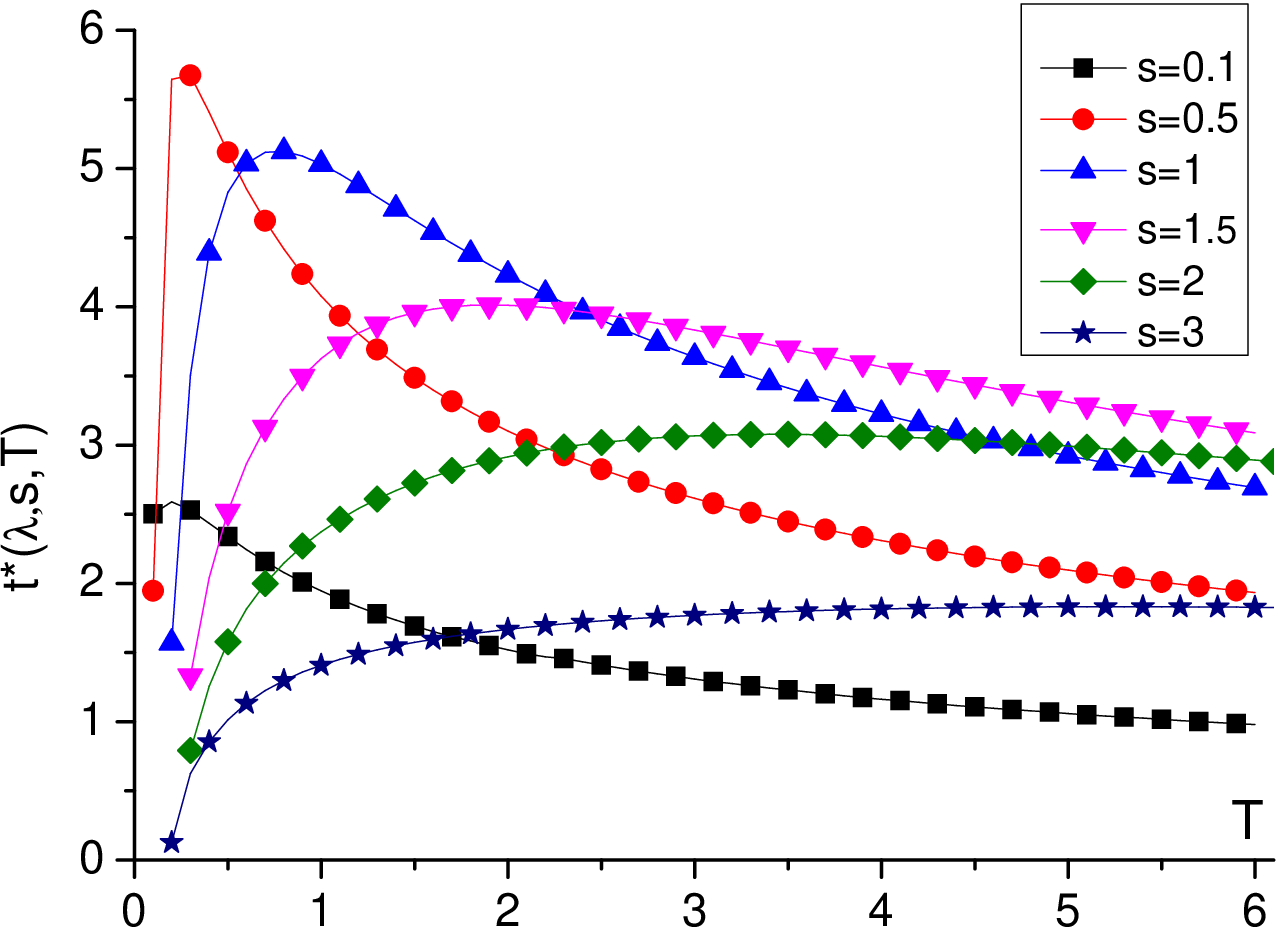}
		\includegraphics[height=0.25\textheight,angle=270]{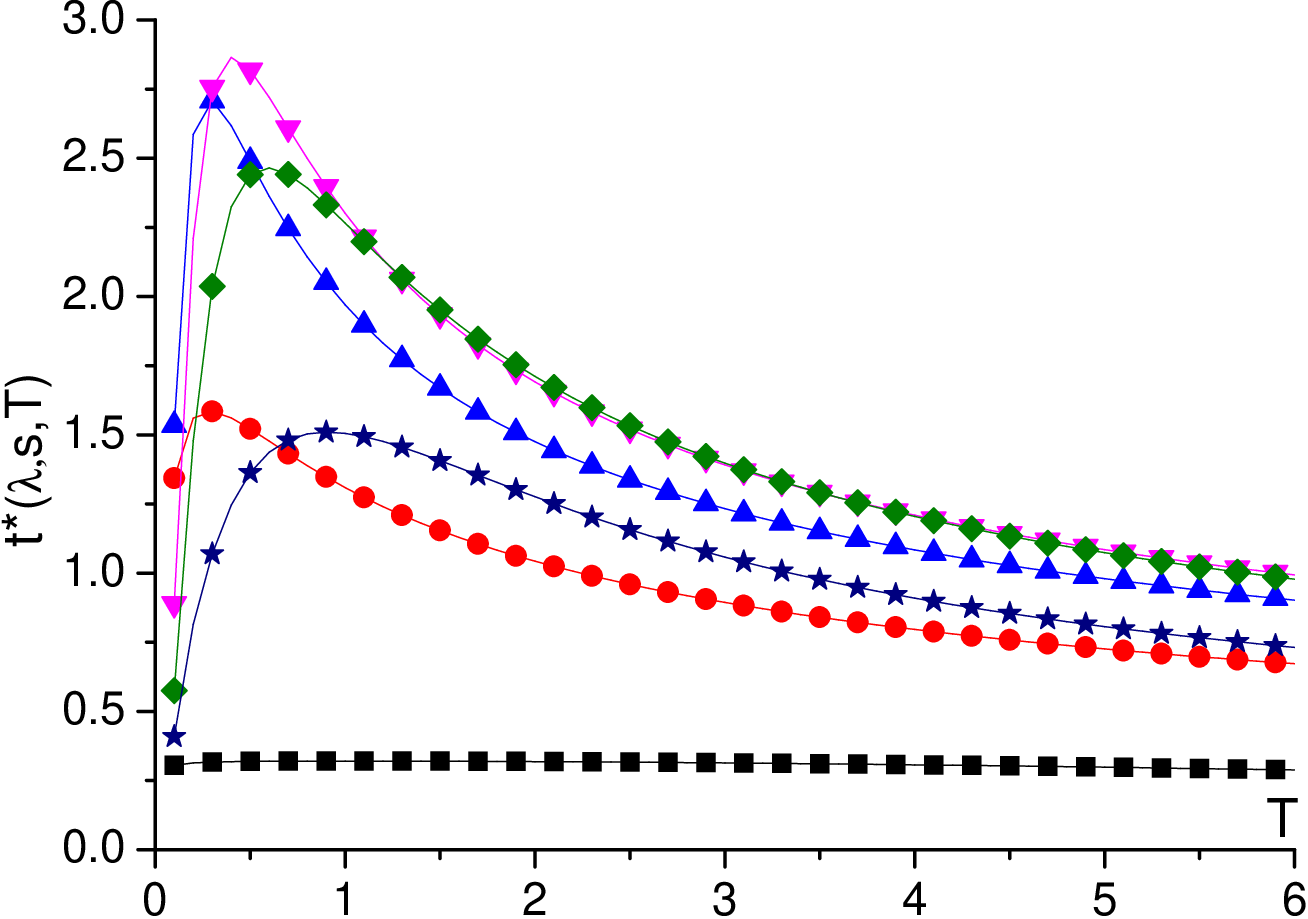}
		\includegraphics[height=0.25\textheight,angle=270]{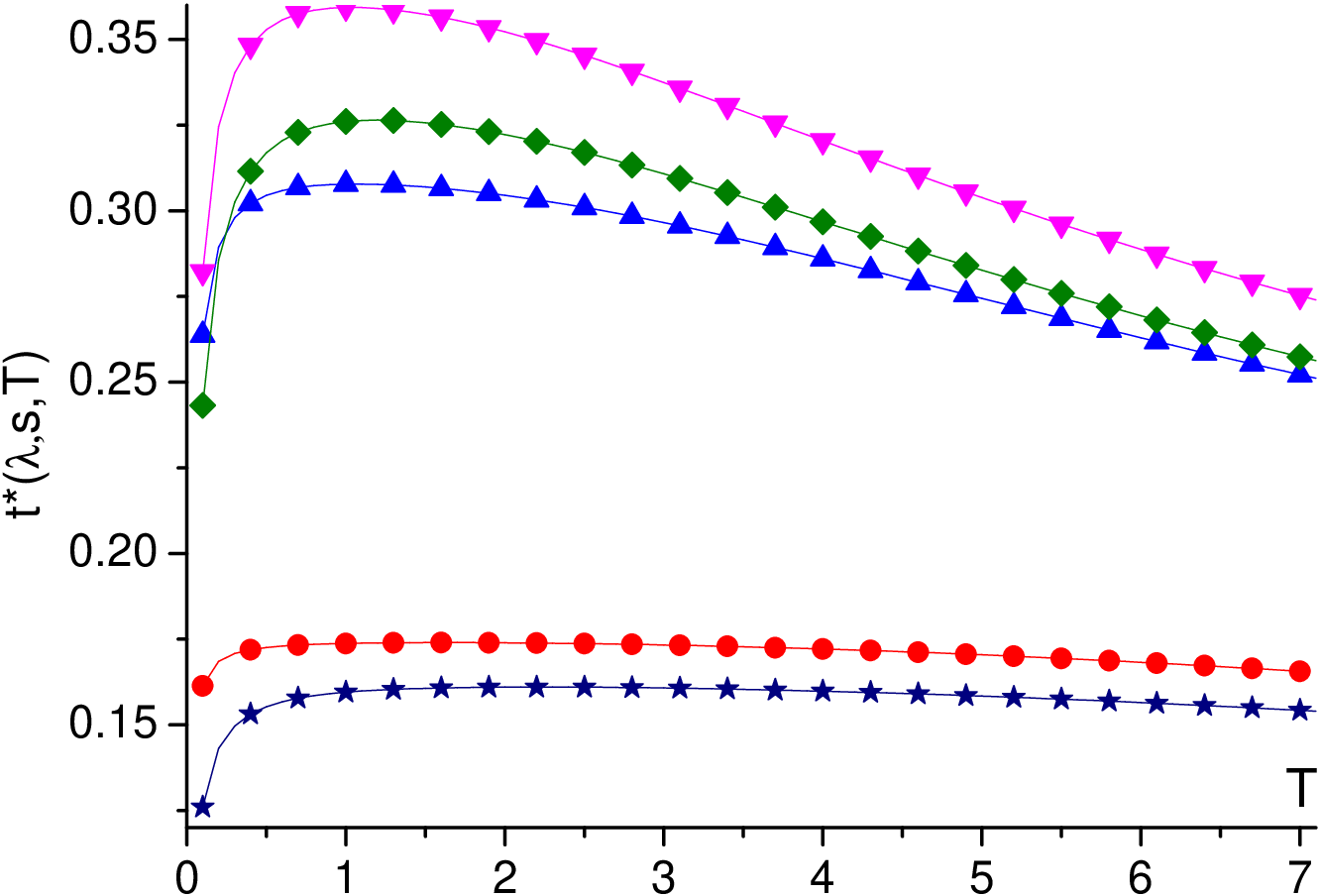}}
	\vspace*{3mm}
	\centerline{	
		\includegraphics[height=0.25\textheight,angle=270]{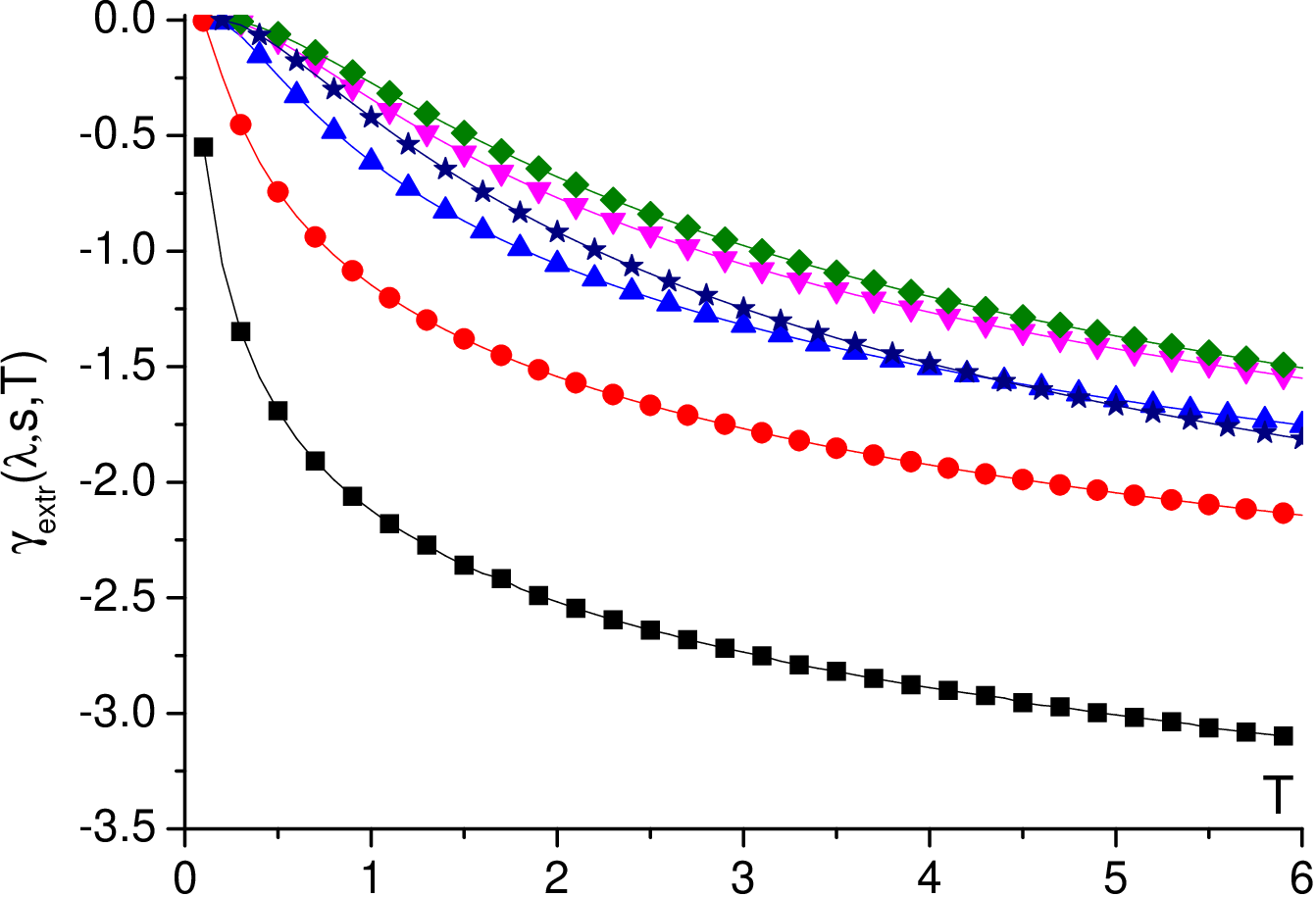}
		\includegraphics[height=0.25\textheight,angle=270]{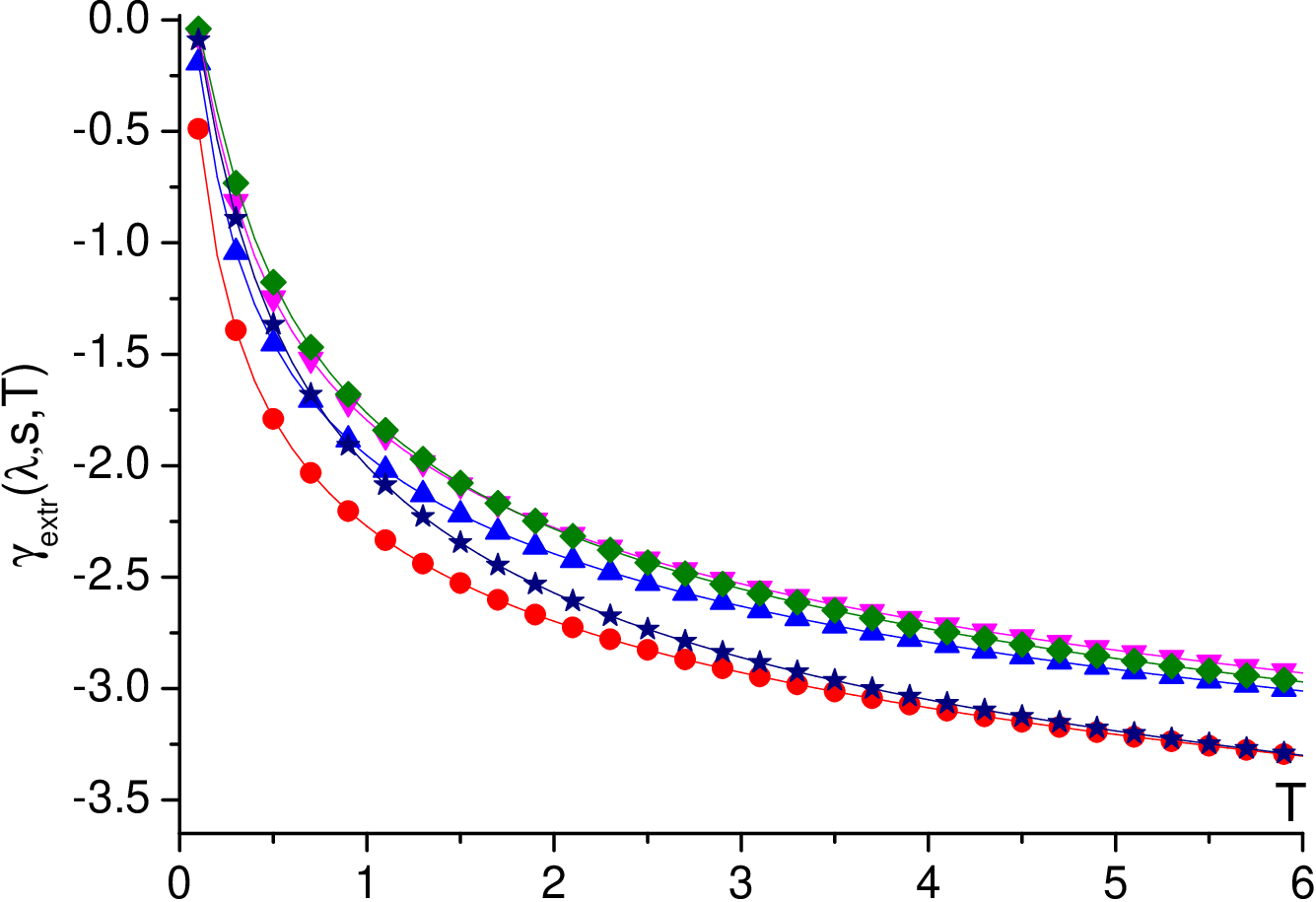}
		\includegraphics[height=0.25\textheight,angle=270]{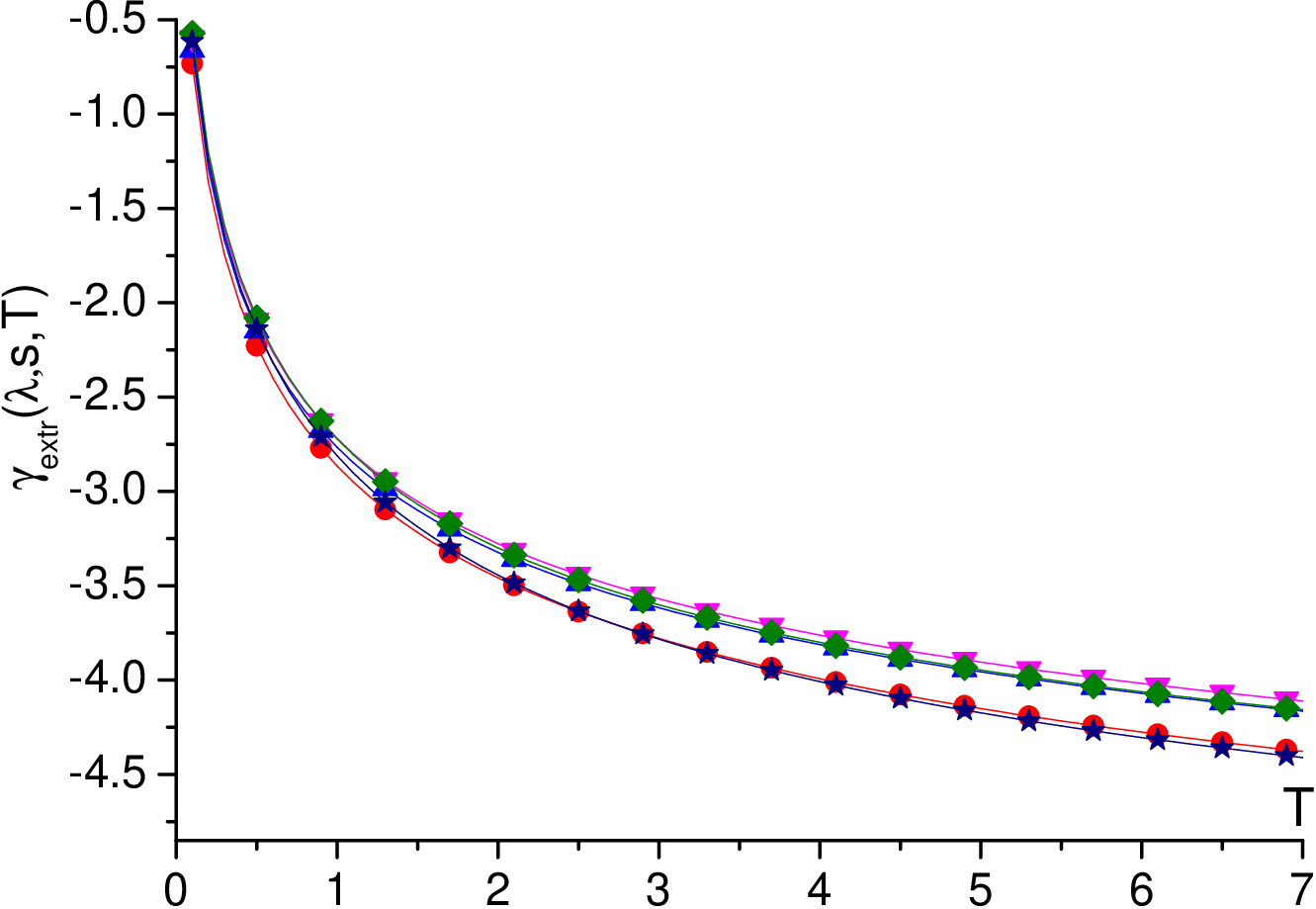}
	}
	\caption{Top row: the recoherence times $t^*(\lambda,s,T)$ in the low (left panel, $\lambda=0.1$), moderate (central panel, $\lambda=1$), and high (right panel,  $\lambda=10$) coupling regimes at different ohmicity indexes $s$ as functions of
	temperature $T$. Bottom row: the same for the minimum values $\gamma_{\rm extr}(\lambda, s,T)$ of the total decoherence function $\gamma_{\rm tot}(t)$.
	}
	\label{tg-T}
\end{figure*}

In Fig.~\ref{tg-s}, we plot the dependence of the recoherence times $t^*(\lambda,s,T)$ and the minimum values $\gamma_{\rm extr}(\lambda, s,T)$ of the total decoherence function $\gamma_{\rm tot}(t)$ on the ohmicity index $s$ at different temperatures starting from $T=0.1$ up to $T=4$. The results are presented for three coupling regimes with interaction strengths $\lambda=0.1$, 1, and 10, respectively. In the advanced sub-Ohmic ($s\ll 1$) and super-Ohmic ($s>5$) regimes, we used the adopted earlier definition of $t^*$: the recoherence times are determined as the 1st intersection point of the corresponding curves with the abscissa axis, see Fig.~\ref{2gamma-oscil}. Hence no $t^*_{\rm tot}$ has been considered. At $0.2<s<4$ there is always a single intersection point, so there is no need to talk about the total recoherence time.

It is seen in Fig.~\ref{tg-s} that the maximum recoherence times are shifted to larger ohmicity indexes with temperature, whereas the minimum values $\gamma_{\rm extr}(\lambda,s,T)$ of the total decoherence function show an opposite tendency. In the high coupling regime, $\gamma_{\rm extr}(\lambda,s,T)$ weakly depends on $s$. 
One can conclude that at the small coupling strength, the maxima of $t^*$ occur mainly in the sub-Ohmic and Ohmic regimes, whereas at the moderate-to-high coupling, the maximum recoherence times are observed in the super-Ohmic regime with $s\sim 3/2$.

It is also essential to emphasize that in the low coupling -- low temperature regime there is quite a broad domain of the ohmicity indexes, where $\gamma_{\rm extr}(\lambda,s,T)\ge 0$, and this is accompanied by a corresponding break of the green curve for $t^*(\lambda,s,T)$. In this domain of ($s-T$) values, the chosen coupling strength $\lambda=0.1$ can be shown to be well below $\lambda_{\rm min}$, see Eqn.~(\ref{lam-min}). At temperatures $T=0.25$ and 0.5, $\lambda_{\rm min}$ is slightly below the chosen coupling strength $\lambda=0.1$; the top points of the corresponding cyan and blue curves for $\gamma_{\rm extr}(\lambda,s,T)$ lie just below zero, and we have a very weak qubit recoherence in this case. 

Note also that the maxima of almost all the curves in the bottom row of Fig.~\ref{tg-s} (except the black one) lie in the $s\sim 2$ region. In this region of the ohmicity indexes, the incomplete decoherence at long times begins to occur \cite{PRA2012}. Therefore, one can suppose that the short-time system recoherence and the long-time dynamics --- the decoherence at large $t$ ---  can sometimes be closely related in such an interesting manner. Namely, the domain of $s$, where the decoherence changes its type (from the complete to incomplete one) is, simultaneously, that of the weakest recoherence. 

One more remark to be made when inspecting Fig.~\ref{tg-s}: the minima $\gamma_{\rm extr}(\lambda,s,T)$ at both small and large $s$ are defined by the expression $-\ln[\coth(\beta\omega_0)/2]$, see also (\ref{oscil}). It is most clearly visible in the moderate and high coupling regimes, where all the curves are starting and terminating at the same horizontal level. 
In the advanced sub- and super-Ohmic regimes, the correlation contribution to the total decoherence function solely defines the dynamics of $\gamma_{\rm tot}(t)$. It is clearly seen in Fig.~\ref{2gamma-oscil}, where during the first recoherence event, the blue and red curves completely coincide and begin to separate only at $i>2$.

A similar behaviour of $t^*$ as function of $T$ (at the fixed ohmicity indexes) can be observed in Fig.~\ref{tg-T}, where the maxima of the recoherence times are shifted with $s$ to higher temperatures. Expectedly, the most pronounced recoherence occurs at high temperatures, as can be seen in the bottom row of Fig.~\ref{tg-T}. Remarkably, the curves for $\gamma_{\rm extr}$ in all the panels are ordered from the bottom to the top, when the ohmicity indexes increase from 0.5 to~2. It means that the sub-Ohmic and Ohmic couplings are more preferable for the recoherence than the super-ohmic ones. However, in the advanced super-Ohmic regime at $s=3$, the corresponding curve (marked by stars) shows an opposite trend and goes below the green and pink symbols.

To conclude this section, let us note that the large values of $\exp[-\gamma_{\rm extr}(\lambda,s,T)] \sim 50\div 70$ in the expression (\ref{coh3}) for coherence, arising at high temperatures (see the bottom row in Fig.~\ref{tg-T}), should not confuse the readers. A considerable enhancement of the coherence with time is always preceded by a large decrease of its initial value,
 as it is clearly seen from Eqs.~(61) and (67) of Ref.~\cite{myPRA2015}. 
Thus, the length of the Bloch vector remains smaller than 1, and the spin system remains in a mixed state, even though its purity can increase significantly. Besides, as follows from the general physical reasoning, low temperatures keep the quantum systems in more coherent (purer) states.  Thus though high temperatures favour an increase of $\gamma_{\rm extr}(\lambda,s,T)$, it is low $T$ that enhance an evident ``purification'' of the system (see lines with blue triangles in Fig.~5 of Ref.~\cite{myPRA2015}).

\section{Conclusions and outlook\label{secV}}

In this paper, we have examined the RDE for a single qubit interacting with a phonon bath within the dephasing model.
The initial state of the composite system (the qubit plus the environment) is prepared
through the non-selective quantum measurements of the special kind, when the Gram operator constructed on the spin state vectors takes a diagonal form. In such a case, the correlation contribution to the total decoherence function 
i) is a non-positive oscillating function, non-linear in the coupling strength, and ii) does not depend on the qubit state.

According to i), when the coupling strengths are large enough, $\gamma_{\rm cor}(t)$ can outweigh the joint influence of the vacuum and thermal fluctuations of the bath, yielding a single (or multiple) RDE. We defined the recoherence time $t^*$ as an instant, when $\gamma_{\rm tot}(t)$ changes its sign from negative to positive, what corresponds to the transition from the coherence enhancement to pure decoherence. We also pay attention to the value of $\gamma_{\rm extr}$, when the total decoherence function becomes minimal. 
We study the dependence of $t^*$ and $\gamma_{\rm extr}$ on the bath characteristics, namely: a) the coupling constant $\lambda$ (which defines an interaction strength between the qubit and its environment), b) the ohmicity parameter $s$ (it describes the nature of interaction), and c) the bath temperature~$T$.

According to ii), we can eliminate the influence of the qubit state on the recoherence dynamics occurring at short times and concentrate our attention solely on the influence of the qubit environment. The qubit state determines only the initial value of the system coherence~\cite{myPRA2015}.

We study the ($s-T)$ dependence of the recoherence times $t^*$ and of $\gamma_{\rm extr}$ for weak, intermediate, and strong coupling. Remarkably, it turns out that the weakest recoherence occurs at $s\sim 2$, where there is a transition from the total to the incomplete decoherence at large times. As a result, the short-time dynamics (the coherence enhancement) and the long-time behaviour (the decoherence) can be interrelated in such an interesting way. Besides, the obtained results give us some hints about the basic characteristics of the bath, which provide the most optimal values of $t^*$ and $\gamma_{\rm extr}$ in some sense. Such research can be quite promising from an applied point of view and deserves to be continued.

We have also calculated dependence of the critical value $\lambda_{\rm min}$ of the coupling strength, which separates the pure decoherence processes (at $\lambda<\lambda_{\rm min}$) from the coherence enhancement (at $\lambda>\lambda_{\rm min}$), on the ohmicity index and the bath temperature. It is interesting to note, that the domain of the parameters, which makes the open quantum system most persistent to the coherence enhancement, to a certain extent overlaps with the characteristics of the environment, which are most favourable to the maximum entanglement of \textit{two spins} in the same dephasing model \cite{entangl2015}. This is another quite interesting interrelation, which combines two quantum effects of a different nature. 

Therefore, it would be interesting to investigate how the eventual coherence enhancement in a two qubit system could be related with the dynamics of the quantum concurrence. There is a common viewpoint \cite{ficel2015} that an increase of the system-environment correlations of the quantum nature favours the entanglement (using an example of the cavity modes interacting with the squeezed reservoir). It would be interesting to find additional evidence supporting or questioning this statement for the case of the correlated thermal bath interacting with two spins. The claims made in Ref.~\cite{ficel2015} that the initial coherence can outlive the initial entanglement, or that the revival of entanglement occurs even after the decoherence, should be verified also for other open quantum systems. In the case studied here, we encountered similar initial spin-environment correlations arising due to the non-selective quantum measurement, which are responsible for the setup of RDE, and, maybe, for the enhancement of the qubits entanglement at short or even at intermediate times. 

We plan to address the above mentioned problems in future.

\bibliography{references}  

\begin{thebibliography}{41}
\expandafter\ifx\csname natexlab\endcsname\relax\def\natexlab#1{#1}\fi
\expandafter\ifx\csname bibnamefont\endcsname\relax
  \def\bibnamefont#1{#1}\fi
\expandafter\ifx\csname bibfnamefont\endcsname\relax
  \def\bibfnamefont#1{#1}\fi
\expandafter\ifx\csname citenamefont\endcsname\relax
  \def\citenamefont#1{#1}\fi
\expandafter\ifx\csname url\endcsname\relax
  \def\url#1{\texttt{#1}}\fi
\expandafter\ifx\csname urlprefix\endcsname\relax\def\urlprefix{URL }\fi
\providecommand{\bibinfo}[2]{#2}
\providecommand{\eprint}[2][]{\url{#2}}

\bibitem[{\citenamefont{Wineland}(2013)}]{Wineland}
\bibinfo{author}{\bibfnamefont{D.~J.} \bibnamefont{Wineland}},
  \bibinfo{journal}{Rev. Mod. Phys.} \textbf{\bibinfo{volume}{85}},
  \bibinfo{pages}{1103} (\bibinfo{year}{2013}),
  \urlprefix\url{https://link.aps.org/doi/10.1103/RevModPhys.85.1103}.

\bibitem[{\citenamefont{Wiseman and Milburn}(2009)}]{q-measurement1}
\bibinfo{author}{\bibfnamefont{H.~M.} \bibnamefont{Wiseman}} \bibnamefont{and}
  \bibinfo{author}{\bibfnamefont{G.~J.} \bibnamefont{Milburn}},
  \emph{\bibinfo{title}{Quantum measurement and control}}
  (\bibinfo{publisher}{Cambridge university press}, \bibinfo{year}{2009}).

\bibitem[{\citenamefont{Saeedi et~al.}(2013)\citenamefont{Saeedi, Simmons,
  Salvail, Dluhy, Riemann, Abrosimov, Becker, Pohl, Morton, and
  Thewalt}}]{39min}
\bibinfo{author}{\bibfnamefont{K.}~\bibnamefont{Saeedi}},
  \bibinfo{author}{\bibfnamefont{S.}~\bibnamefont{Simmons}},
  \bibinfo{author}{\bibfnamefont{J.~Z.} \bibnamefont{Salvail}},
  \bibinfo{author}{\bibfnamefont{P.}~\bibnamefont{Dluhy}},
  \bibinfo{author}{\bibfnamefont{H.}~\bibnamefont{Riemann}},
  \bibinfo{author}{\bibfnamefont{N.~V.} \bibnamefont{Abrosimov}},
  \bibinfo{author}{\bibfnamefont{P.}~\bibnamefont{Becker}},
  \bibinfo{author}{\bibfnamefont{H.-J.} \bibnamefont{Pohl}},
  \bibinfo{author}{\bibfnamefont{J.~J.~L.} \bibnamefont{Morton}},
  \bibnamefont{and} \bibinfo{author}{\bibfnamefont{M.~L.~W.}
  \bibnamefont{Thewalt}}, \bibinfo{journal}{Science}
  \textbf{\bibinfo{volume}{342}}, \bibinfo{pages}{830} (\bibinfo{year}{2013}),
  \eprint{https://www.science.org/doi/pdf/10.1126/science.1239584},
  \urlprefix\url{https://www.science.org/doi/abs/10.1126/science.1239584}.

\bibitem[{\citenamefont{Ignatyuk and Morozov}(2015)}]{myPRA2015}
\bibinfo{author}{\bibfnamefont{V.~V.} \bibnamefont{Ignatyuk}} \bibnamefont{and}
  \bibinfo{author}{\bibfnamefont{V.~G.} \bibnamefont{Morozov}},
  \bibinfo{journal}{Phys. Rev. A} \textbf{\bibinfo{volume}{91}},
  \bibinfo{pages}{052102} (\bibinfo{year}{2015}),
  \urlprefix\url{https://link.aps.org/doi/10.1103/PhysRevA.91.052102}.

\bibitem[{\citenamefont{Zhang et~al.}(2015)\citenamefont{Zhang, Han, Xia, Yu,
  and Fan}}]{trapping}
\bibinfo{author}{\bibfnamefont{Y.-J.} \bibnamefont{Zhang}},
  \bibinfo{author}{\bibfnamefont{W.}~\bibnamefont{Han}},
  \bibinfo{author}{\bibfnamefont{Y.-J.} \bibnamefont{Xia}},
  \bibinfo{author}{\bibfnamefont{Y.-M.} \bibnamefont{Yu}}, \bibnamefont{and}
  \bibinfo{author}{\bibfnamefont{H.}~\bibnamefont{Fan}},
  \bibinfo{journal}{Scientific reports} \textbf{\bibinfo{volume}{5}},
  \bibinfo{pages}{13359} (\bibinfo{year}{2015}).

\bibitem[{\citenamefont{Lastra et~al.}(2010)\citenamefont{Lastra, Reyes, and
  Wallentowitz}}]{charge-q-bit}
\bibinfo{author}{\bibfnamefont{F.}~\bibnamefont{Lastra}},
  \bibinfo{author}{\bibfnamefont{S.~A.} \bibnamefont{Reyes}}, \bibnamefont{and}
  \bibinfo{author}{\bibfnamefont{S.}~\bibnamefont{Wallentowitz}},
  \bibinfo{journal}{Journal of Physics B: Atomic, Molecular and Optical
  Physics} \textbf{\bibinfo{volume}{44}}, \bibinfo{pages}{015504}
  (\bibinfo{year}{2010}),
  \urlprefix\url{https://dx.doi.org/10.1088/0953-4075/44/1/015504}.

\bibitem[{\citenamefont{Hsiang and Ford}(2009)}]{Hsiang2009}
\bibinfo{author}{\bibfnamefont{J.-T.} \bibnamefont{Hsiang}} \bibnamefont{and}
  \bibinfo{author}{\bibfnamefont{L.~H.} \bibnamefont{Ford}},
  \bibinfo{journal}{International Journal of Modern Physics A}
  \textbf{\bibinfo{volume}{24}}, \bibinfo{pages}{1705} (\bibinfo{year}{2009}),
  \eprint{https://doi.org/10.1142/S0217751X09045273},
  \urlprefix\url{https://doi.org/10.1142/S0217751X09045273}.

\bibitem[{\citenamefont{Chin et~al.}(2013)\citenamefont{Chin, Prior, Rosenbach,
  Caycedo-Soler, Huelga, and Plenio}}]{pp-complexes}
\bibinfo{author}{\bibfnamefont{A.~W.} \bibnamefont{Chin}},
  \bibinfo{author}{\bibfnamefont{J.}~\bibnamefont{Prior}},
  \bibinfo{author}{\bibfnamefont{R.}~\bibnamefont{Rosenbach}},
  \bibinfo{author}{\bibfnamefont{F.}~\bibnamefont{Caycedo-Soler}},
  \bibinfo{author}{\bibfnamefont{S.~F.} \bibnamefont{Huelga}},
  \bibnamefont{and} \bibinfo{author}{\bibfnamefont{M.~B.}
  \bibnamefont{Plenio}}, \bibinfo{journal}{Nature Physics}
  \textbf{\bibinfo{volume}{9}}, \bibinfo{pages}{113} (\bibinfo{year}{2013}).

\bibitem[{\citenamefont{Chisholm et~al.}(2021)\citenamefont{Chisholm,
  Garc{\'i}a-P{\'e}rez, Rossi, Maniscalco, and Palma}}]{chisholm2021witnessing}
\bibinfo{author}{\bibfnamefont{D.~A.} \bibnamefont{Chisholm}},
  \bibinfo{author}{\bibfnamefont{G.}~\bibnamefont{Garc{\'i}a-P{\'e}rez}},
  \bibinfo{author}{\bibfnamefont{M.~A.~C.} \bibnamefont{Rossi}},
  \bibinfo{author}{\bibfnamefont{S.}~\bibnamefont{Maniscalco}},
  \bibnamefont{and} \bibinfo{author}{\bibfnamefont{G.~M.} \bibnamefont{Palma}},
  \bibinfo{journal}{Quantum Science and Technology}
  \textbf{\bibinfo{volume}{7}}, \bibinfo{pages}{015022} (\bibinfo{year}{2021}),
  \urlprefix\url{https://dx.doi.org/10.1088/2058-9565/ac40f3}.

\bibitem[{\citenamefont{Macchiavello et~al.}(2000)\citenamefont{Macchiavello,
  Palma, and Zeilinger}}]{macchiavello2000quantum}
\bibinfo{author}{\bibfnamefont{C.}~\bibnamefont{Macchiavello}},
  \bibinfo{author}{\bibfnamefont{G.}~\bibnamefont{Palma}}, \bibnamefont{and}
  \bibinfo{author}{\bibfnamefont{A.}~\bibnamefont{Zeilinger}},
  \emph{\bibinfo{title}{Quantum Computation and Quantum Information Theory:
  Reprint Volume with Introductory Notes for ISI TMR Network School, 12-23 July
  1999, Villa Gualino, Torino, Italy}}, G - Reference,Information and
  Interdisciplinary Subjects Series (\bibinfo{publisher}{World Scientific},
  \bibinfo{year}{2000}), ISBN \bibinfo{isbn}{9789810241179},
  \urlprefix\url{https://books.google.com.ua/books?id=t3BqDQAAQBAJ}.

\bibitem[{\citenamefont{Reina et~al.}(2002)\citenamefont{Reina, Quiroga, and
  Johnson}}]{reina2002}
\bibinfo{author}{\bibfnamefont{J.~H.} \bibnamefont{Reina}},
  \bibinfo{author}{\bibfnamefont{L.}~\bibnamefont{Quiroga}}, \bibnamefont{and}
  \bibinfo{author}{\bibfnamefont{N.~F.} \bibnamefont{Johnson}},
  \bibinfo{journal}{Phys. Rev. A} \textbf{\bibinfo{volume}{65}},
  \bibinfo{pages}{032326} (\bibinfo{year}{2002}),
  \urlprefix\url{https://link.aps.org/doi/10.1103/PhysRevA.65.032326}.

\bibitem[{\citenamefont{Haikka et~al.}(2013)\citenamefont{Haikka, Johnson, and
  Maniscalco}}]{haikka2013non}
\bibinfo{author}{\bibfnamefont{P.}~\bibnamefont{Haikka}},
  \bibinfo{author}{\bibfnamefont{T.~H.} \bibnamefont{Johnson}},
  \bibnamefont{and}
  \bibinfo{author}{\bibfnamefont{S.}~\bibnamefont{Maniscalco}},
  \bibinfo{journal}{Phys. Rev. A} \textbf{\bibinfo{volume}{87}},
  \bibinfo{pages}{010103} (\bibinfo{year}{2013}),
  \urlprefix\url{https://link.aps.org/doi/10.1103/PhysRevA.87.010103}.

\bibitem[{\citenamefont{Addis et~al.}(2014)\citenamefont{Addis, Bylicka,
  Chru\ifmmode \acute{s}\else \'{s}\fi{}ci\ifmmode~\acute{n}\else
  \'{n}\fi{}ski, and Maniscalco}}]{nonMark-1}
\bibinfo{author}{\bibfnamefont{C.}~\bibnamefont{Addis}},
  \bibinfo{author}{\bibfnamefont{B.}~\bibnamefont{Bylicka}},
  \bibinfo{author}{\bibfnamefont{D.}~\bibnamefont{Chru\ifmmode \acute{s}\else
  \'{s}\fi{}ci\ifmmode~\acute{n}\else \'{n}\fi{}ski}}, \bibnamefont{and}
  \bibinfo{author}{\bibfnamefont{S.}~\bibnamefont{Maniscalco}},
  \bibinfo{journal}{Phys. Rev. A} \textbf{\bibinfo{volume}{90}},
  \bibinfo{pages}{052103} (\bibinfo{year}{2014}),
  \urlprefix\url{https://link.aps.org/doi/10.1103/PhysRevA.90.052103}.

\bibitem[{\citenamefont{Budini}(2018)}]{budini}
\bibinfo{author}{\bibfnamefont{A.~A.} \bibnamefont{Budini}},
  \bibinfo{journal}{Phys. Rev. A} \textbf{\bibinfo{volume}{97}},
  \bibinfo{pages}{052133} (\bibinfo{year}{2018}),
  \urlprefix\url{https://link.aps.org/doi/10.1103/PhysRevA.97.052133}.

\bibitem[{\citenamefont{Lombardo and Villar}(2015)}]{lombardo2015}
\bibinfo{author}{\bibfnamefont{F.~C.} \bibnamefont{Lombardo}} \bibnamefont{and}
  \bibinfo{author}{\bibfnamefont{P.~I.} \bibnamefont{Villar}},
  \bibinfo{journal}{Phys. Rev. A} \textbf{\bibinfo{volume}{91}},
  \bibinfo{pages}{042111} (\bibinfo{year}{2015}),
  \urlprefix\url{https://link.aps.org/doi/10.1103/PhysRevA.91.042111}.

\bibitem[{\citenamefont{Giraldi}(2017)}]{giraldi2017}
\bibinfo{author}{\bibfnamefont{F.}~\bibnamefont{Giraldi}},
  \bibinfo{journal}{Phys. Rev. A} \textbf{\bibinfo{volume}{95}},
  \bibinfo{pages}{022109} (\bibinfo{year}{2017}),
  \urlprefix\url{https://link.aps.org/doi/10.1103/PhysRevA.95.022109}.

\bibitem[{\citenamefont{Clos and Breuer}(2012)}]{clos2012}
\bibinfo{author}{\bibfnamefont{G.}~\bibnamefont{Clos}} \bibnamefont{and}
  \bibinfo{author}{\bibfnamefont{H.-P.} \bibnamefont{Breuer}},
  \bibinfo{journal}{Phys. Rev. A} \textbf{\bibinfo{volume}{86}},
  \bibinfo{pages}{012115} (\bibinfo{year}{2012}),
  \urlprefix\url{https://link.aps.org/doi/10.1103/PhysRevA.86.012115}.

\bibitem[{\citenamefont{Morozov et~al.}(2012)\citenamefont{Morozov, Mathey, and
  R\"opke}}]{PRA2012}
\bibinfo{author}{\bibfnamefont{V.~G.} \bibnamefont{Morozov}},
  \bibinfo{author}{\bibfnamefont{S.}~\bibnamefont{Mathey}}, \bibnamefont{and}
  \bibinfo{author}{\bibfnamefont{G.}~\bibnamefont{R\"opke}},
  \bibinfo{journal}{Phys. Rev. A} \textbf{\bibinfo{volume}{85}},
  \bibinfo{pages}{022101} (\bibinfo{year}{2012}),
  \urlprefix\url{https://link.aps.org/doi/10.1103/PhysRevA.85.022101}.

\bibitem[{\citenamefont{Breuer et~al.}(2009)\citenamefont{Breuer, Laine, and
  Piilo}}]{gamma-minus-1}
\bibinfo{author}{\bibfnamefont{H.-P.} \bibnamefont{Breuer}},
  \bibinfo{author}{\bibfnamefont{E.-M.} \bibnamefont{Laine}}, \bibnamefont{and}
  \bibinfo{author}{\bibfnamefont{J.}~\bibnamefont{Piilo}},
  \bibinfo{journal}{Phys. Rev. Lett.} \textbf{\bibinfo{volume}{103}},
  \bibinfo{pages}{210401} (\bibinfo{year}{2009}),
  \urlprefix\url{https://link.aps.org/doi/10.1103/PhysRevLett.103.210401}.

\bibitem[{\citenamefont{Vacchini and Breuer}(2010)}]{gamma-minus-2}
\bibinfo{author}{\bibfnamefont{B.}~\bibnamefont{Vacchini}} \bibnamefont{and}
  \bibinfo{author}{\bibfnamefont{H.-P.} \bibnamefont{Breuer}},
  \bibinfo{journal}{Phys. Rev. A} \textbf{\bibinfo{volume}{81}},
  \bibinfo{pages}{042103} (\bibinfo{year}{2010}),
  \urlprefix\url{https://link.aps.org/doi/10.1103/PhysRevA.81.042103}.

\bibitem[{\citenamefont{Tang et~al.}(2012)\citenamefont{Tang, Li, Li, Zou, Guo,
  Breuer, Laine, and Piilo}}]{gamma-minus-3}
\bibinfo{author}{\bibfnamefont{J.-S.} \bibnamefont{Tang}},
  \bibinfo{author}{\bibfnamefont{C.-F.} \bibnamefont{Li}},
  \bibinfo{author}{\bibfnamefont{Y.-L.} \bibnamefont{Li}},
  \bibinfo{author}{\bibfnamefont{X.-B.} \bibnamefont{Zou}},
  \bibinfo{author}{\bibfnamefont{G.-C.} \bibnamefont{Guo}},
  \bibinfo{author}{\bibfnamefont{H.-P.} \bibnamefont{Breuer}},
  \bibinfo{author}{\bibfnamefont{E.-M.} \bibnamefont{Laine}}, \bibnamefont{and}
  \bibinfo{author}{\bibfnamefont{J.}~\bibnamefont{Piilo}},
  \bibinfo{journal}{Europhysics Letters} \textbf{\bibinfo{volume}{97}},
  \bibinfo{pages}{10002} (\bibinfo{year}{2012}),
  \urlprefix\url{https://dx.doi.org/10.1209/0295-5075/97/10002}.

\bibitem[{\citenamefont{Breuer et~al.}(2016)\citenamefont{Breuer, Laine, Piilo,
  and Vacchini}}]{colloquium2016}
\bibinfo{author}{\bibfnamefont{H.-P.} \bibnamefont{Breuer}},
  \bibinfo{author}{\bibfnamefont{E.-M.} \bibnamefont{Laine}},
  \bibinfo{author}{\bibfnamefont{J.}~\bibnamefont{Piilo}}, \bibnamefont{and}
  \bibinfo{author}{\bibfnamefont{B.}~\bibnamefont{Vacchini}},
  \bibinfo{journal}{Rev. Mod. Phys.} \textbf{\bibinfo{volume}{88}},
  \bibinfo{pages}{021002} (\bibinfo{year}{2016}),
  \urlprefix\url{https://link.aps.org/doi/10.1103/RevModPhys.88.021002}.

\bibitem[{\citenamefont{Rivas et~al.}(2014)\citenamefont{Rivas, Huelga, and
  Plenio}}]{rivas2014}
\bibinfo{author}{\bibfnamefont{{\'A}.}~\bibnamefont{Rivas}},
  \bibinfo{author}{\bibfnamefont{S.~F.} \bibnamefont{Huelga}},
  \bibnamefont{and} \bibinfo{author}{\bibfnamefont{M.~B.}
  \bibnamefont{Plenio}}, \bibinfo{journal}{Reports on Progress in Physics}
  \textbf{\bibinfo{volume}{77}}, \bibinfo{pages}{094001}
  (\bibinfo{year}{2014}),
  \urlprefix\url{https://dx.doi.org/10.1088/0034-4885/77/9/094001}.

\bibitem[{\citenamefont{Dajka et~al.}(2013)\citenamefont{Dajka, Gardas, and
  {\L}uczka}}]{luczka2013}
\bibinfo{author}{\bibfnamefont{J.}~\bibnamefont{Dajka}},
  \bibinfo{author}{\bibfnamefont{B.}~\bibnamefont{Gardas}}, \bibnamefont{and}
  \bibinfo{author}{\bibfnamefont{J.}~\bibnamefont{{\L}uczka}},
  \bibinfo{journal}{International Journal of Theoretical Physics}
  \textbf{\bibinfo{volume}{52}}, \bibinfo{pages}{1148} (\bibinfo{year}{2013}).

\bibitem[{\citenamefont{Breuer and Petruccione}(2007)}]{BPbook}
\bibinfo{author}{\bibfnamefont{H.-P.} \bibnamefont{Breuer}} \bibnamefont{and}
  \bibinfo{author}{\bibfnamefont{F.}~\bibnamefont{Petruccione}},
  \emph{\bibinfo{title}{The Theory of Open Quantum Systems}}
  (\bibinfo{publisher}{Oxford University Press}, \bibinfo{year}{2007}), ISBN
  \bibinfo{isbn}{9780199213900},
  \urlprefix\url{https://doi.org/10.1093/acprof:oso/9780199213900.001.0001}.

\bibitem[{\citenamefont{Chaudhry and Gong}(2013)}]{chaudhry2013}
\bibinfo{author}{\bibfnamefont{A.~Z.} \bibnamefont{Chaudhry}} \bibnamefont{and}
  \bibinfo{author}{\bibfnamefont{J.}~\bibnamefont{Gong}},
  \bibinfo{journal}{Phys. Rev. A} \textbf{\bibinfo{volume}{87}},
  \bibinfo{pages}{012129} (\bibinfo{year}{2013}),
  \urlprefix\url{https://link.aps.org/doi/10.1103/PhysRevA.87.012129}.

\bibitem[{\citenamefont{Ignatyuk}(2015)}]{overcomplete}
\bibinfo{author}{\bibfnamefont{V.~V.} \bibnamefont{Ignatyuk}},
  \bibinfo{journal}{Phys. Rev. A} \textbf{\bibinfo{volume}{92}},
  \bibinfo{pages}{062115} (\bibinfo{year}{2015}),
  \urlprefix\url{https://link.aps.org/doi/10.1103/PhysRevA.92.062115}.

\bibitem[{\citenamefont{Giraldi}(2021)}]{giraldi2021}
\bibinfo{author}{\bibfnamefont{F.}~\bibnamefont{Giraldi}},
  \bibinfo{journal}{Open Systems \& Information Dynamics}
  \textbf{\bibinfo{volume}{28}}, \bibinfo{pages}{2150002}
  (\bibinfo{year}{2021}), \eprint{https://doi.org/10.1142/S1230161221500025},
  \urlprefix\url{https://doi.org/10.1142/S1230161221500025}.

\bibitem[{\citenamefont{Giraldi}(2022)}]{giraldi2022short}
\bibinfo{author}{\bibfnamefont{F.}~\bibnamefont{Giraldi}},
  \bibinfo{journal}{International Journal of Quantum Information}
  \textbf{\bibinfo{volume}{20}}, \bibinfo{pages}{2250003}
  (\bibinfo{year}{2022}), \eprint{https://doi.org/10.1142/S0219749922500034},
  \urlprefix\url{https://doi.org/10.1142/S0219749922500034}.

\bibitem[{\citenamefont{Braginsky et~al.}(1992)\citenamefont{Braginsky,
  Khalili, and Thorne}}]{q-measurement}
\bibinfo{author}{\bibfnamefont{V.~B.} \bibnamefont{Braginsky}},
  \bibinfo{author}{\bibfnamefont{F.~Y.} \bibnamefont{Khalili}},
  \bibnamefont{and} \bibinfo{author}{\bibfnamefont{K.~S.}
  \bibnamefont{Thorne}}, \emph{\bibinfo{title}{Quantum Measurement}}
  (\bibinfo{publisher}{Cambridge University Press}, \bibinfo{year}{1992}).

\bibitem[{\citenamefont{Kraus}(1983)}]{kraus}
\bibinfo{author}{\bibfnamefont{K.}~\bibnamefont{Kraus}},
  \bibinfo{journal}{States, Effects and Operations. Fundamental Notions of
  Quantum Theory} \textbf{\bibinfo{volume}{190}} (\bibinfo{year}{1983}).

\bibitem[{\citenamefont{Holevo}(2001)}]{Holevo2001}
\bibinfo{author}{\bibfnamefont{A.~S.} \bibnamefont{Holevo}},
  \emph{\bibinfo{title}{Statistical structure of quantum theory}}
  (\bibinfo{publisher}{Springer Science \& Business Media},
  \bibinfo{year}{2001}).

\bibitem[{\citenamefont{Huang}(2005)}]{SRM}
\bibinfo{author}{\bibfnamefont{S.}~\bibnamefont{Huang}},
  \bibinfo{journal}{Phys. Rev. A} \textbf{\bibinfo{volume}{72}},
  \bibinfo{pages}{022324} (\bibinfo{year}{2005}),
  \urlprefix\url{https://link.aps.org/doi/10.1103/PhysRevA.72.022324}.

\bibitem[{\citenamefont{Unruh}(1995)}]{Unruh}
\bibinfo{author}{\bibfnamefont{W.~G.} \bibnamefont{Unruh}},
  \bibinfo{journal}{Phys. Rev. A} \textbf{\bibinfo{volume}{51}},
  \bibinfo{pages}{992} (\bibinfo{year}{1995}),
  \urlprefix\url{https://link.aps.org/doi/10.1103/PhysRevA.51.992}.

\bibitem[{\citenamefont{Morozov and R{\"o}pke}(2012)}]{MR-CMP2012}
\bibinfo{author}{\bibfnamefont{V.}~\bibnamefont{Morozov}} \bibnamefont{and}
  \bibinfo{author}{\bibfnamefont{G.}~\bibnamefont{R{\"o}pke}},
  \bibinfo{journal}{Condensed Matter Physics} \textbf{\bibinfo{volume}{15}},
  \bibinfo{pages}{43004} (\bibinfo{year}{2012}),
  \urlprefix\url{https://icmp.lviv.ua/journal/zbirnyk.72/43004/art43004.pdf}.

\bibitem[{\citenamefont{Ignatyuk and Morozov}(2013)}]{myCMP}
\bibinfo{author}{\bibfnamefont{V.}~\bibnamefont{Ignatyuk}} \bibnamefont{and}
  \bibinfo{author}{\bibfnamefont{V.}~\bibnamefont{Morozov}},
  \bibinfo{journal}{Condensed Matter Physics} \textbf{\bibinfo{volume}{16}},
  \bibinfo{pages}{34001} (\bibinfo{year}{2013}),
  \urlprefix\url{https://icmp.lviv.ua/journal/zbirnyk.75/34001/art34001.pdf}.

\bibitem[{\citenamefont{Leggett et~al.}(1987)\citenamefont{Leggett,
  Chakravarty, Dorsey, Fisher, Garg, and Zwerger}}]{Leggett}
\bibinfo{author}{\bibfnamefont{A.~J.} \bibnamefont{Leggett}},
  \bibinfo{author}{\bibfnamefont{S.}~\bibnamefont{Chakravarty}},
  \bibinfo{author}{\bibfnamefont{A.~T.} \bibnamefont{Dorsey}},
  \bibinfo{author}{\bibfnamefont{M.~P.~A.} \bibnamefont{Fisher}},
  \bibinfo{author}{\bibfnamefont{A.}~\bibnamefont{Garg}}, \bibnamefont{and}
  \bibinfo{author}{\bibfnamefont{W.}~\bibnamefont{Zwerger}},
  \bibinfo{journal}{Rev. Mod. Phys.} \textbf{\bibinfo{volume}{59}},
  \bibinfo{pages}{1} (\bibinfo{year}{1987}),
  \urlprefix\url{https://link.aps.org/doi/10.1103/RevModPhys.59.1}.

\bibitem[{\citenamefont{Tan et~al.}(2015)\citenamefont{Tan, Chaudhry, and
  Gong}}]{entangl2015}
\bibinfo{author}{\bibfnamefont{D.~Y.} \bibnamefont{Tan}},
  \bibinfo{author}{\bibfnamefont{A.~Z.} \bibnamefont{Chaudhry}},
  \bibnamefont{and} \bibinfo{author}{\bibfnamefont{J.}~\bibnamefont{Gong}},
  \bibinfo{journal}{Journal of Physics B: Atomic, Molecular and Optical
  Physics} \textbf{\bibinfo{volume}{48}}, \bibinfo{pages}{115505}
  (\bibinfo{year}{2015}),
  \urlprefix\url{https://dx.doi.org/10.1088/0953-4075/48/11/115505}.

\bibitem[{\citenamefont{Hill and Wootters}(1997)}]{Wooters1}
\bibinfo{author}{\bibfnamefont{S.~A.} \bibnamefont{Hill}} \bibnamefont{and}
  \bibinfo{author}{\bibfnamefont{W.~K.} \bibnamefont{Wootters}},
  \bibinfo{journal}{Phys. Rev. Lett.} \textbf{\bibinfo{volume}{78}},
  \bibinfo{pages}{5022} (\bibinfo{year}{1997}),
  \urlprefix\url{https://link.aps.org/doi/10.1103/PhysRevLett.78.5022}.

\bibitem[{\citenamefont{Wootters}(1998)}]{Wooters2}
\bibinfo{author}{\bibfnamefont{W.~K.} \bibnamefont{Wootters}},
  \bibinfo{journal}{Phys. Rev. Lett.} \textbf{\bibinfo{volume}{80}},
  \bibinfo{pages}{2245} (\bibinfo{year}{1998}),
  \urlprefix\url{https://link.aps.org/doi/10.1103/PhysRevLett.80.2245}.

\bibitem[{\citenamefont{Aloufi et~al.}(2015)\citenamefont{Aloufi, Bougouffa,
  and Ficek}}]{ficel2015}
\bibinfo{author}{\bibfnamefont{K.}~\bibnamefont{Aloufi}},
  \bibinfo{author}{\bibfnamefont{S.}~\bibnamefont{Bougouffa}},
  \bibnamefont{and} \bibinfo{author}{\bibfnamefont{Z.}~\bibnamefont{Ficek}},
  \bibinfo{journal}{Physica Scripta} \textbf{\bibinfo{volume}{90}},
  \bibinfo{pages}{074020} (\bibinfo{year}{2015}),
  \urlprefix\url{https://dx.doi.org/10.1088/0031-8949/90/7/074020}.

\end{thebibliography}

\end{document}